\documentclass[nonumber]{aa}

\makeatletter
\@ifpackageloaded{lineno}{\nolinenumbers}{}
\makeatother

\usepackage{graphicx}
\usepackage{txfonts}
\usepackage{hyperref}
\usepackage{lastpage}
\hypersetup{colorlinks = true, citecolor = {blue}, urlcolor= {blue}}

\def\gapprox{\;\rlap{\lower 3.0pt                       
        \hbox{$\sim$}}\raise 2.5pt\hbox{$>$}\;}
\def\lapprox{\;\rlap{\lower 3.1pt                       
        \hbox{$\sim$}}\raise 2.7pt\hbox{$<$}\;}

\newcommand{\be}{ \begin{equation} }
\newcommand{\ee}{\end{equation}}

\newcommand{\ben}{\begin{enumerate}}
\newcommand{\een}{\end{enumerate}}

\renewenvironment{thebibliography}[1]{\begin{oldthebibliography}{#1}\setlength{\parskip}{0ex}\setlength{\itemsep}{0ex}}{\end{oldthebibliography}}

\usepackage{diagbox}
\usepackage{siunitx}
\newcommand{\orcid}[1]{\href{https://orcid.org/#1}{\protect\includegraphics[width=8pt]{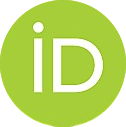}}}
\makeatletter
\renewcommand*\aa@pageof{, page \thepage{} of \pageref*{LastPage}}
\makeatother

\usepackage[dvipsnames]{xcolor}

\definecolor{darkgreen}{RGB}{31, 207, 31}

\begin{document}

\title{Globular clusters of the Gaia Enceladus/Sausage}
\subtitle{I. Orbital and dynamical evolution on cosmological timescales}

\author{
Mykyta~Bilodid
\inst{1}\orcid{0009-0009-9104-6015},
Maryna~Ishchenko
\inst{1,2,3}\orcid{0000-0002-6961-8170},
Peter~Berczik
\inst{2,3,1}\orcid{0000-0003-4176-152X}
}

\institute{Main Astronomical Observatory,                    National 
           Academy of Sciences of Ukraine,
           27 Akademika Zabolotnoho St, 03143 Kyiv, Ukraine  \email{\href{mailto:bilodid_m@mao.kiev.ua}{bilodid\_m@mao.kiev.ua}}
           \and
            Nicolaus Copernicus Astronomical Centre Polish Academy of Sciences, ul. Bartycka 18, 00-716 Warsaw, Poland
            \and
           Fesenkov Astrophysical Institute, Observatory 23, 050020 Almaty, Kazakhstan
           }
   
\date{Received xxx / Accepted xxx}

\abstract
{The history of our Galaxy is shaped by significant merger events, which contribute to its mass and to the distribution of stars, but which also bring globular clusters that act as the main tracers of the accretion history of the Milky Way.}
{We investigated Gaia-Enceladus/Sausage globular cluster samples and studied their orbital and dynamical evolution over cosmological timescales in external time-variable potential. We estimated the limits of distribution of the escaped stars from the globular clusters' orbital evolution in energy angular momentum space.}
{To reconstruct the orbital evolution of the known globular clusters of the dwarf galaxy Gaia-Enceladus/Sausage, we used the parallel $N$-body code $\varphi$-GPU. We investigated the relationship between globular clusters and their progenitor by analysing their orbital parameters and phase-space distribution during 9 Gyr of evolution in the past. We created a $N$-body model of Gaia-Enceladus/Sausage globular clusters and analysed their dynamical evolution and distribution of the escaped stars today.}
{We summarised the samples of the Gaia-Enceladus/Sausage globular clusters and created two main categories: `most probable' and `tentative', with 15 and 9 clusters, respectively. We analysed the evolution of their kinematic, orbital, and phase-space parameters in the external time-variable potential. We defined phase-space distribution limits of stars that escape from globular clusters during 9 Gyr of evolution: a specific energy from -18 to -12.2 $\times10^4$ km$^2$ s$^{-2}$, L$_{\rm z}$ from -0.98 to 0.72 $\times10^3$ kpc km s$^{-1}$, and L$_{\rm perp}$ from 0 to 1.8 $\times10^3$ kpc km s$^{-1}$. The limits of the GE/S debris in Galactic area based on orbital parameters of the GC's escaped stars are: for apocentre and pericetre distances of 10--28 and 1--4 kpc, < 18 kpc in Galactocentric radius and < |15| kpc in the Z direction. Generally we compared the phase-space distribution of escaped stars from the GCs GE/S debris energy-angular momentum limits with the observed very metal-poor stars, which belong to the GE/S itself and produce consistent results.}
{}


\titlerunning{GCs of the GE/S: evolution}
\authorrunning{M.~Bilodid et al.}
\maketitle

\section{Introduction}\label{sec:Intr}

One of the rapidly developing fields of research in astronomy today is Galactic archaeology (\citealt{Frebel2010, Miglio2013, Spitoni2020, Koppelman2020, Xiang2022, Deason2024}), which studies the history of the assembly of the Milky Way (MW). Based on observational data \citep{Helmi2020, Buder2021, Vallenari2023}, both astrometric and spectroscopic, as well as numerical simulations \citep{Wang2024, Merrow2024}. Galactic archaeology investigates the evolutionary processes that shaped our Galaxy from the earliest stages of the Universe's existence, including a series of mergers between the MW and its neighbours.

Today we are able to identify at least five major merger events: Sagittarius, Gaia Enceladus (often referred to as Sausage, GE/S), Sequoia, Helmi, and Kraken, which are key to understanding the MW's current structure and dynamics \citep{Belokurov2006, Helmi2018, Myeong2019,  Ibata2021, Kruijssen2020, Matsuno2022, Mateu2023}. Each of these events not only brought stars and dark matter but also contributed whole globular cluster (GC) systems. These GCs are today identified as an ex situ GC population in the MW. They are vital for tracing the global dynamical history of the Galaxy \citep{Monkman2006, Lind2015, Escudero2018, Giusti2024}. Their long lifetimes \citep{Vandenberg1996, VandenBerg2013, Valcin2021} and stable internal structure \citep{Ishchenko2024massloss} make them excellent candidates for studying past interactions and large merger events, providing clues about the types of galaxies that contributed to the growth of MW and the processes involved in these mergers. Moreover, analysis of the chemical compositions and dynamics of these GCs can shed light on the different conditions and environments of their formation, enriching our understanding of galactic evolution. This framework of using GCs as a tracers highlights the intricate and often violent dynamical assembling history of our Galaxy, illustrating how it has evolved from a collection of smaller systems into the large spiral disc galaxy as we see today.

According to cosmological simulations, there is a correlation between massive merger events and possible highly radial orbits of satellite galaxies \citep{Fattahi2019, Mackereth2019}. The accretion process of large stellar systems, such as GE/S, into the host galaxy's potential is a non-trivial task. This process leads to a complex distribution of energy and angular momentum in the debris of the accreted galaxy \citep{Quinn1984, Quinn1986, Tormen1998, Bosch1999}, as well as to a complex chemical gradient \citep{Kirby2011, Ho2015} in the Galaxy.

For instance, it is believed that the local stellar disc of the MW was primarily formed as a result of a merger with the massive GE/S satellite dwarf galaxy 8--11 Gyr ago or z $\sim$1--2 \citep{Helmi2018, Belokurov2018, Pfeffer2020}. This dwarf galaxy had a mass of $\sim$10$^{8-9.5}$ M$_\odot$, accounting for $\sim$10\% of the MW’s mass at that time. This event was one of the most significant mergers in the MW and the last major one \citep{Ruchti2015, Fragkoudi2020}. Additionally, it has been suggested that the GE/S merger also might have played a significant role in the formation of the MW's bar; see \cite{Kruijssen2009}. 

In the study of \cite{Koppelman2020} from the set of extensive $N$-body simulations we can conclude, that the most comparable results for the {\it Gaia} DR2 velocity distribution of the MW solar vicinity (< 1 kpc) halo stars we obtain for the disc merger on the retrograde orbit with the orbital inclination angle of $\sim$30 degree. In addition, more than 20 GCs were added to the MW as a result of this merger \citep{Massari2019, Kruijssen2020, Malhan2022, Chen2024, CARMA2025a, CARMA2025b}. These clusters are crucial for studying the Galaxy’s past interactions and dynamics. The merger also triggered intense bursts of star formation, contributing to the thickening of the MW’s disc. 

As a first step in our current investigation, we carried out an extensive literature search for possible GE/S related GC systems. Here we mainly focused on the studies of \cite{CARMA2025b}, \cite{Chen2024}, \cite{Malhan2022}, \cite{Kruijssen2020}, and \cite{Massari2019}, which cover all the main aspects of observational and theoretical backgrounds of the MW GCs phase-space distribution analyses.

Based on the {\it Gaia} DR3 data for this set of GCs, we ran a 9 Gyr lookback time orbital simulation in the time-variable, cosmologically motivated, IllustrisTNG selected halo potential \citep{Ishchenko2023a}. Using this kinematical model for GCs orbital evolution and applying the energy and angular momentum phase-space limits for the GE/S objects from the above literature, we created our own list of the most probable GCs with a strong GE/S progenitor connection. 

As a final step, we modelled, using the high-order dynamical $N$-body code, our set of probable GE/S GCs forward dynamical evolution (from -9 Gyr) up until today. Based on the detail orbital integration and stellar density distribution study of the escaped stars from these GCs, we finally redefined the current GE/S stream phase-space boundaries using our external time-variable potential (TVP). We also compared our GC's escaped stars distribution with the catalogues of the very metal-poor (VMP) stars, observed today in the GE/S debris. Based on our calculations of the external galaxy model, we redefined the boundaries of the GE/S stream itself. 

The paper is organised as follows. In Sect.~\ref{sec:data-sample} we analyse GCs, collected from the literature which may belong to the GE/S dwarf galaxy. In Sect.~\ref{sec:orb-reconstr} we reconstruct the orbits of selected GCs up to a lookback time of 9 Gyr in a time-variable external galactic potential. In Sect.~\ref{sec:compar}, we prepare a sample of GCs associated with the GE/S as a progenitor, based on the evolution of the orbital parameters. In Sects.~\ref{sec:n-body-model} and~\ref{sec:compar-vpm}, we carry out an $N$-body dynamical simulation of the selected GCs and compare the distribution of their escaped stars with the observed stars in the GE/S debris in energy and angular momentum phase-space.

\section{Data sample}\label{sec:data-sample}

In this section, we give an overview of the literature and analyse the different authors' classification prescription of GCs associated with the GE/S dwarf galaxy. The \textit{Gaia} space telescopes have contributed significantly to this process. Thanks to the last catalogues of DR2 and DR3 \cite{Gaia2018, Gaia2020, Vallenari2023} of \textit{Gaia}, it has become possible to combine two methods of classification: kinematics and chemical. The present study employs a comparative approach that integrates the kinematic method with the analysis of chemical patterns. The results offer a comprehensive exploration of the interplay between GC orbital parameters and chemical patterns. In the following, we summarise the classification of the GC membership to the GE/S based on recent publications.

In earlier pioneering investigations, authors \cite{Mackey2004} and \cite{Recio-Blanco2018} suggested that $\sim$ 50 GCs may be formed ex situ and fall onto to our Galaxy’s gravitational potential during approximately seven to eight continuous mergers together with their dwarf galaxies. In \cite{Kruijssen2019}, the authors make the suggestion that our Galaxy might have undergone at last three major mergers with different dwarf galaxies. The authors also make a sample of GCs that are associated with the remnants of the Canis Major, Kraken, and Sagittarius dwarf galaxies. This sample is based on summarising information about metal abundances from earlier works, and comparing it in the age-metallicity space. We need to note that most of the clusters from their sample of Canis Major GCs are identified in later works as systems that belong to the GE/S dwarf galaxy. In \cite{Myeong2018}, the authors used integrals of motion to separate GCs into different groups. In a later publication (\cite{Myeong2019}), the authors find 21 potential GE/S GCs based on their own dynamical analysis. 

Recently \cite{CARMA2025a, CARMA2025b} determined the age and metallicity of 13 GCs that are associated with the GE/S dwarf galaxy, applying the data from the HST telescope and theoretical isochron fitting, based on the new {\tt BaSTI} stellar evolution model mesh \citep{BaSTI2018}. For the association of the selected GCs with the GE/S stream, the authors applied only their chemical analyses. The obtained age-metallicity relationship separates the clusters into two formation epochs, with a time interval of about 2 Gyr. As a final result, the authors present the list of GCs potentially associated with the GE/S stream: NGC 288, NGC 362, NGC 1261, NGC 1851, NGC 2298, NGC 2808, NGC 5286, NGC 5897, NGC 6205, NGC 6341, NGC 6779, NGC 7089, and NGC 7099.

In the study of \cite{Chen2024}, the authors proposed a method of classifying progenitors of GCs based on the analysis of kinematic characteristics (i.e. integrals of motion and action angles) together with their chemical composition and age. The authors applied their own developed clustering method using up to ten different parameters, including orbital parameters (actions $J_r$, $J_{\varphi}$, and $J_{\theta}$, apocenter r$_{apo}$ and pericenter distances r$_{peri}$, eccentricities, {\tt ecc}, Galactocentric distance, $D$, and total energy, $E$) with their chemical properties (age and metallicity). It needs to be noted that for the kinematic analysis the authors applied the `static' galactic potential based on the standard three-component model of \cite{Bovy2015}. The authors associated the next GCs with the GE/S dwarf: ESO-SC06, IC 1257, NGC 1261, NGC 1851, NGC 1904, NGC 2298, NGC 2808, NGC 288, NGC 362, NGC 4147, NGC 5286, NGC 6205, NGC 6229, NGC 6341, NGC 6779, NGC 6864, NGC 7089, NGC 7099, NGC 7492, Palomar 2, NGC 5634, NGC 5904, NGC 6981, NGC 7078, NGC 6426, and NGC 6584.

In \cite{Malhan2022}, the authors proposed the classification of $\sim$170 GCs that combined their chemical and kinematic parameters. The authors used extensive sky surveys (LAMOST DR7 and APOGEE DR17) together with astrometric data collected from the \textit{Gaia} EDR3 catalogue. The classification was carried out by calculating action angles together for GCs and progenitors (streams). They applied the static and axisymmetric model of the Galactic potential \citep{McMillan2017}. The final list of GCs that \cite{Malhan2022} attribute to the GE/S dwarf galaxy is: NGC 7492, NGC 6229, NGC 6584, NGC 5634, IC 1257, NGC 1851, NGC 2298, NGC 4147, NGC 1261, NGC 6981, NGC 1904, NGC 7089, NGC 5904, and NGC 6864.

In earlier papers by \cite{Massari2019} and \cite{Kruijssen2020} on the determination of the GCs membership with the GE/S, the authors mainly used astrometric data that was taken from the earlier \textit{Gaia} DR2. During their investigations, they applied the same static axisymmetric Galactic potential \citep{McMillan2017} to compare the orbital parameters of the GCs and the GE/S stream. 
The list of GCs that \cite{Massari2019} attribute to the GE/S dwarf galaxy is: Djorg 1, NGC 4833, NGC 5897, NGC 6235, NGC 6284, Terzan 10, NGC 5139, ESO-SC06, IC 1257, NGC 1261, NGC 1851, NGC 1904, NGC 2298, NGC 2808, NGC 288, NGC 362, NGC 4147, NGC 5286, NGC 6205, NGC 6229, NGC 6341, NGC 6779, NGC 6864, NGC 7089, NGC 7099, NGC 7492, Palomar 2, NGC 5634, NGC 5904, Palomar 15, NGC 6101, and NGC 3201. 

Contrary to the investigation above, in the work of \cite{Kruijssen2020} only the {\tt ecc} and r$_{apo}$ parameters as we see them today were studied. Here the authors also applied the simple age-metallicity estimation for GCs. The resulting list of GCs is attributed to the GE/S dwarf galaxy: NGC 288, NGC 362, NGC 1261, NGC 1851, NGC 1904, NGC 2298, NGC 2808, NGC 4147, NGC 4833, NGC 5286, NGC 5897, NGC 6205, NGC 6235, NGC 6284, NGC 6341, NGC 6779, NGC 6864, NGC 7089, NGC 7099, NGC 7492, and NGC 5139. 

In Table~\ref{tab:gc-pre-sets} we present the combined sample of GCs associated with the GE/S, based on the described above publications. Our table consists of several pre-sets of GCs. For the first pre-set, we assumed GCs that we see in all five publications (4 GCs). For the second pre-set, we combined the GCs that appeared in at least four papers (12 GCs). The other two pre-sets have objects from the lists of only three, two, or even one publication. Based on this, we assume that these sets possibly contain the less probable (or less confident) GCs from the GE/S stream members.

\begin{table}[htbp]
\caption{Pre-sets of the GCs that could potentially originate from the GE/S dwarf galaxy.}
\centering
\sisetup{separate-uncertainty}
\resizebox{0.50\textwidth}{!}{
\begin{tabular}{c|c|c}
\hline
\hline
 Sampling          &  Paper & GC's name \\
              &          &           \\
\hline
\hline 
\raisebox{-1.5ex}[0pt][0pt]{\centering pre-set-1}
                 & A-B-C-D-E     & NGC 1261, NGC 1851,  NGC 2298    \\
                 &               & NGC 7089     \\

\hline
\raisebox{-5.5ex}[0pt][0pt]{\centering pre-set-2}
        & B-C-D-E       &  NGC 1904, NGC 4147, NGC 5634    \\
        &               &  NGC 5904, NGC 6864, NGC 7492     \\
        &               &  NGC 2808     \\
        & A-C-D-E       &  NGC 362, NGC 5286, NGC 6341   \\
        &               &  NGC 6779, NGC 7099    \\

\hline
\raisebox{-3ex}[0pt][0pt]{\centering pre-set-3}
        & B-C-E         & IC 1257, NGC 6229   \\
        & B-D-E         & NGC 288, NGC 6205     \\
        & D-E           & NGC 5139, NGC 5897 \\
\hline
\raisebox{-7ex}[0pt][0pt]{\centering pre-set-4}
        & B-C           & NGC 6981, NGC 6584                    \\
        & B-E           & ESO-SC06, Pal 2    \\
        & D-E           & NGC 3201, NGC 4833,  NGC 6101 \\
        &               & NGC 6235, NGC 6284 \\
        & B             & NGC 7078, NGC 6426\\
        & E             & Djorg1, Terzan10, Pal 15\\
\hline
\hline
\end{tabular}}
\tablefoot{\cite{CARMA2025b} as A, \cite{Chen2024} as B, \cite{Malhan2022} as C, \cite{Kruijssen2020} as D, \cite{Massari2019} as E.}
\label{tab:gc-pre-sets}
\end{table} 

\section{Collected GCs' orbital reconstruction as a point mass in time-variating potential} \label{sec:orb-reconstr}

In this section, we carry out the orbital reconstruction of 36 GCs from the Table~\ref{tab:gc-pre-sets} inside the Galactic time-variable external potential. This gives us the opportunity to investigate the orbital evolution inside the MW of our selected GCs on the cosmological timescale. 

For orbital reconstruction and numerical integration, we used the high-order parallel \textit{N}-body code $\varphi$-GPU, which is based on fourth-order Hermite integration together with the hierarchical individual block timestep scheme \cite{Berczik2011, BSW2013}. This code was already successfully applied  to the MW GCs global dynamical evolution in our previous works \cite{Ishchenko2023a, Ishchenko2023b, Ishchenko2023c}. To reconstruct orbits and calculate the main orbital parameters for our pre-sets of the 36 GCs during 9 Gyr of lookback evolution, we integrated each of them as a single point mass. The current positions, proper motions, radial velocities, and heliocentric distances were selected from the catalogue of  \cite{Baumgardt2021} \footnote{\url{https://people.smp.uq.edu.au/HolgerBaumgardt/globular/orbits.html}} and presented in Table~\ref{tab:cur-pos}.

To calculate the Cartesian Galactocentric co-ordinates, we used the Sun position $X_{\odot} = -8.178$ ~kpc, $Y_{\odot} = 0$ ~kpc, and $Z_{\odot} = 0.0208$ ~kpc \cite{Gravity2019,Bennett2019}. For the local standard of rest (LSR) velocity, we used $234.737$ ~km s$^{-1}$ \cite{Reid2004}, and for the peculiar velocity of the Sun with respect to the LSR, we used $U_{\odot} = 11.1$ ~km s$^{-1}$, $V_{\odot} = 12.24$ ~km s$^{-1}$, and $W_{\odot} = 7.25$ ~km s$^{-1}$, \citep{Schonrich2010}. In detail the co-ordinate and velocity transformations are described in \cite{Johnson1987}. For the equatorial position of the north galactic pole (NGP), we used the updated values from \cite{Karim2017}: RA$_{\rm NGP} = 192^{\circ}.7278$, DEC$_{\rm NGP} = 26^{\circ}.8630$, and $\theta_{\rm NGP} = 122^{\circ}.9280$.  

\begin{figure}[]
\centering
\includegraphics[width=0.99\linewidth]{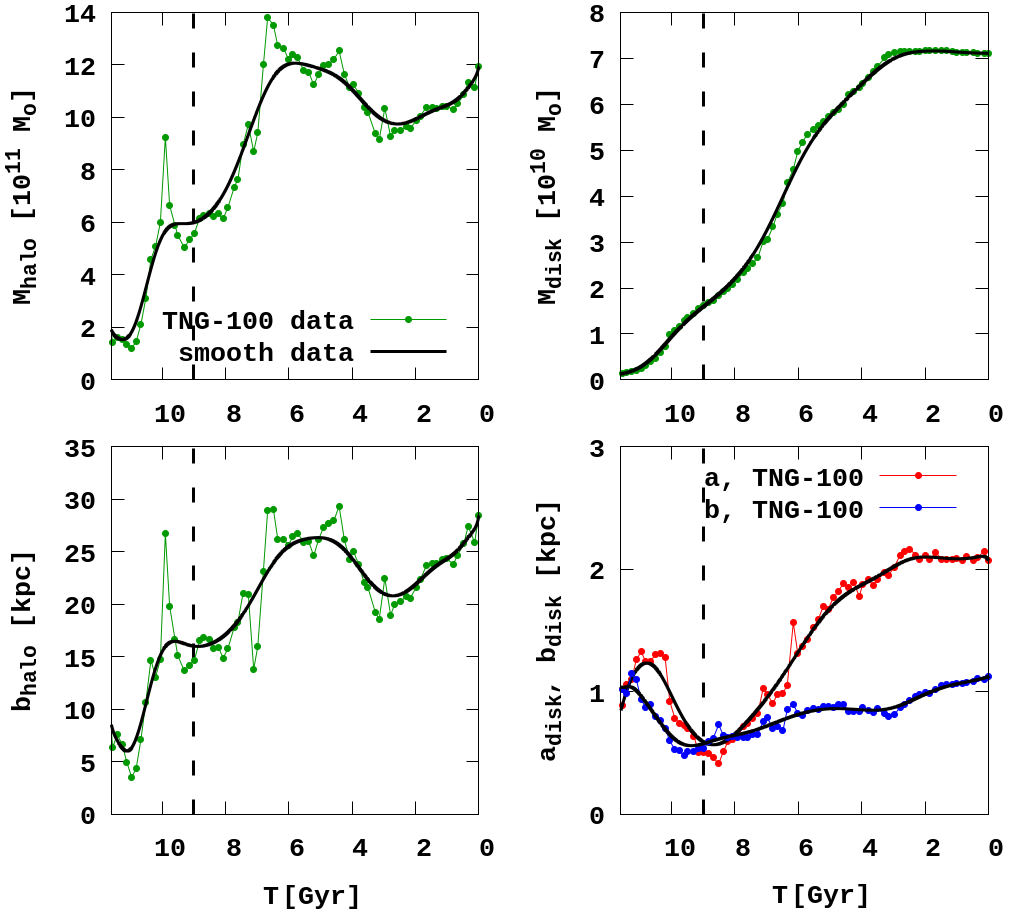}
\caption{Evolution of halo and disc masses, and their characteristic scales for {\tt 411321} external potential that are shown as thick black lines. Lines with dots represent the original TNG-100 data. The dashed line represents the 9 Gyr.}
\label{fig:ext-pot}
\end{figure}

For our purpose we applied the time-variable external MW potential \cite{Ishchenko2023a}, based on the IllustrisTNG-100 cosmological simulation database \cite{Nelson2019,NelsonPill2019}. From our pre-selected list of the MW-like potentials, we used the {\tt 411321} TVP presented in Fig.~\ref{fig:ext-pot}. This potential at present has global parameters very similar to our Galaxy today (for more details, see Sect. 2.2 in \citealt{Ishchenko2023a}). The potential consists of two components: a dark matter halo, described by the potential, $\Phi_{halo}$, from Eq.~\ref{eq.pot} of \cite{Navarro1997}, and a disc, described by the potential, $\Phi_{disc}$, from Eq.~\ref{eq.pot} of \cite{Miyamoto1975}:
\begin{equation}
\begin{aligned}
\Phi=\Phi_{halo}+\Phi_{disc}= \\
- \frac{G M_{halo} \ln \left(1+\frac{\sqrt{R^2+z^2}}{b_{halo}}\right)}{\sqrt{R^2+z^2}} - \frac{G M_{disc}}{\sqrt{R^2+\left(a_{disc}+\sqrt{z^2+b_{disc}^2}\right)^2}}
\label{eq.pot}  
\end{aligned}
.\end{equation}
In the above equation, $R=\sqrt{x^2+y^2}$ is the radial distance from the Galactic center, $M_{halo}$ and $M_{disc}$ are the mass of the MW halo and the mass of the MW disc, respectively, and $b_{halo}$, $a_{disc}$, and $ b_{disc}$ are the spatial scales of the halo and disc potentials. Figure~\ref{fig:ext-pot} shows the time evolution of the main TVP potential {\tt 411321} parameters $M_{halo}$, $M_{disc}$, $b_{halo}$, $a_{disc}$, and $b_{disc}$ during the whole 12 Gyr time evolution of the MW system. Recently, our TNG TVP potentials (particularly the {\tt 411321} model) were also included in the very popular \textit{N}-body Galactic dynamical code {\tt PeTar} \citep{LWPeTar2020}. One can find the recent implementation of the updated {\tt PeTar} code in a project GitHub link~\footnote{\url{https://github.com/lwang-astro/PeTar}}. 

\begin{figure*}[htbp!]
\centering
\includegraphics[width=0.85\linewidth]{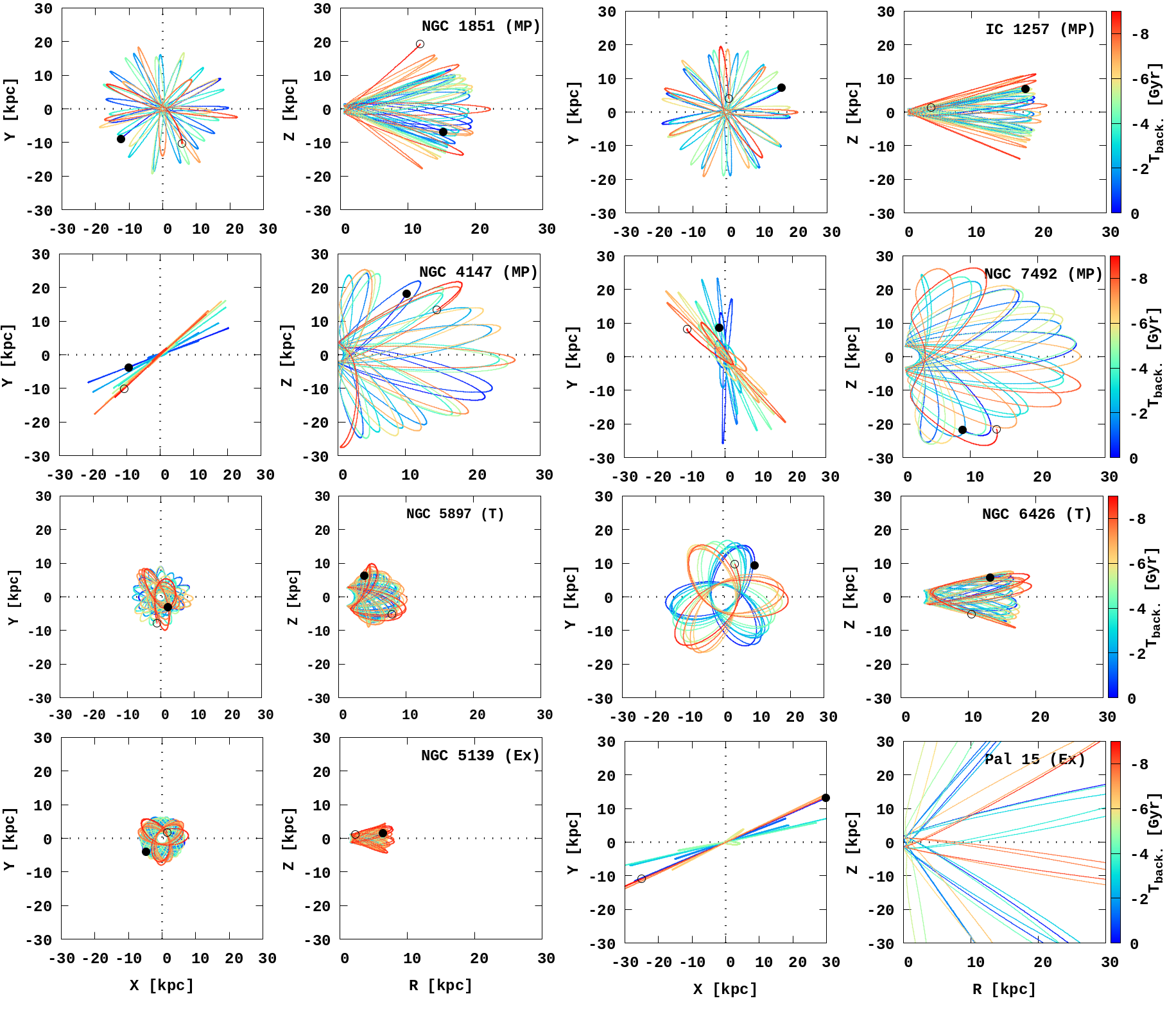}
\caption{Orbital evolution in the {\tt 411321} TVP external potential, presented in $X-Y$ an $R-Z$, where $R$ is the planar Galactocentric radius. The total time of integration is a 9 Gyr lookback time which is represented by the coloured bar. The filled black circle shows the current position, non-filled -- at -9 Gyr ago.}
\label{fig:orb-plots}
\end{figure*}

As a illustration of our GCs typical orbits, we present in Fig.~\ref{fig:orb-plots}, the 9 Gyr backward orbital evolution of the selected clusters from our list (see more detail explanation in the Sec.~\ref{subsec:diss}). As we see for example orbits in our TNG TVP potential, they have a significant long term orbital evolution including the strong precession of the orbits along the Galactic $Z$ axis. 

Also, in Table~\ref{tab:cur-pos} we show the relative errors for each GCs in proper motions, radial velocities, and heliocentric distances based on \textit{Gaia} DR3 measurements \citep{Baumgardt2021}. Most of the GCs have high-accuracy measurements in the 6D kinematic parameters, with relative errors below $\sim$2\%. In some specific cases, due to, the low proper motion of NGC 6584, the error in $\mu_\alpha$ above $\sim$10\%. For the NGC 7089 cluster radial velocity error, we go up to $\sim$5\%. The detailed analyses about errors are represented in Appendix~\ref{sec:gaia-errors}. In Appendix~\ref{sec:gaia-errors} we also discuss the error values of 6D components for each GC and their possible influence on the long-term orbital evolution. 

\section{Procedure of identification GCs as GE/S membership in TVP}\label{sec:compar}

Here, we describe the (i) concept of identifying GCs as possible members of the GE/S stream based on (ii) energy--angular momentum evolution and orbital parameters in a TVP. In subsequent sections, we present (iii) our final sample of GCs connected with the GE/S, classified in this way together with the discussion (iv) about uncertainty surrounding the origin of some of these GCs.

\subsection{Concept of identification}\label{subsec:concept}

Table~\ref{tab:ges-GC-today} shows the possible limits in the phase-space parameters of GCs with GE/S connection, which we collected from recent publications \citep{Sun2023, Malhan2022, Massari2019, Limberg2022}. In works \cite{Sun2023}, \cite{Massari2019}, and \cite{Malhan2022}, the authors mainly use the current distribution in phase-space of kinematic parameters and integrals of motion for the selected GCs. Based on these criteria,  they define the possible ranges for the association of concrete GCs with the GE/S stream. In article \cite{Limberg2022} the authors set the limits by distribution of possible individual GE/S stellar sample that have been selected by their chemical abundances. The authors made a cleaner GE/S sample by discarding stars with chemical abundances compatible with the in situ regions of the [Al/Fe] versus [Mg/Mn] space. In contrast to our current investigation, all of these previous studies of course only defined the current dynamical parameters, i.e. they did not study the possible changes in GCs and GE/S stream orbital motion in the past. 

As we see from Table~\ref{tab:ges-GC-today}, the limits (or boundaries) for L$_z$, total energy, and {\tt ecc} are defined in most of the studies. The boundaries for L$_{perp}$ and r$_{apo}$ are defined in only a few of these studies. In Figs.~\ref{fig:E-lz-sets} and ~\ref{fig:a-e-sets}, we highlighted these parameter boundaries as different colour areas. One can note that the total energy range is quite different between these publications. The \cite{Malhan2022} study especially has a quite narrow energy range, as does the GCs from the GE/S stream members. 

In the previous section, we describe our $N$-body modelling with the aim of orbital motion reconstruction of 36 GCs inside {\tt 411321} TVP up to 9 Gyr lookback time. We show the reconstructed orbital evolution for each GC (colour thin lines), grouped according to the pre-sets from Table~\ref{tab:gc-pre-sets}.   

We assume that GCs can have a GE/S dwarf galaxy origin if more than 80\% of their trajectories during the full 9 Gyr evolution fall within the selected coloured areas; see Figs.~\ref{fig:E-lz-sets} and ~\ref{fig:a-e-sets}. These areas represent the parameters for L$_z$ and L$_{perp}$, together with the total energy as the three main identification criteria. The second criterion is the orbital parameters $a$, {\tt ecc}, r$_{apo}$, and r$_{peri}$. In cases of disputed identification, whether or not a GC belongs to GE/S, chemical patterns will play a decisive role.

\subsection{GC orbital parameters  matching GE/S limits}\label{subsec:compar-e-ltot}

\begin{figure*}[htbp!]
\centering
\includegraphics[width=0.95\linewidth]{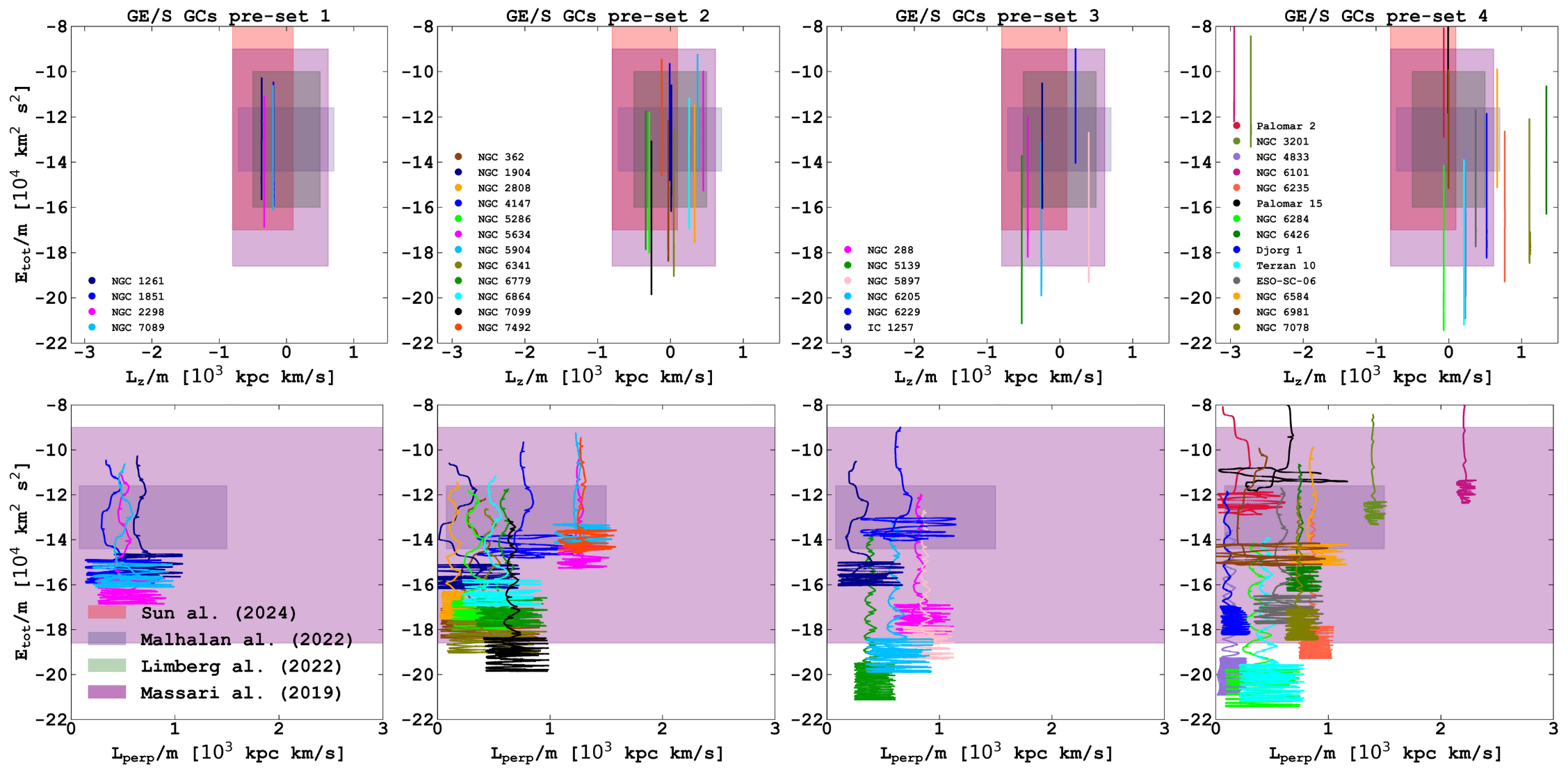}
\caption{Evolution of the GCs phase-space parameters during the 9 Gyr lookback time integration in {\tt 411321} TVP external potential for the four different pre-sets of GE/S GC's (see Table \ref{tab:gc-pre-sets}). The upper row shows the evolution of the L$_{z}$ and total energy; the bottom row of L$_{perp}$ and total energy. All values are in specific units. Transparent coloured areas represent the limits of the phase-space for GE/S GCs based on publications from Table~\ref{tab:ges-GC-today}.}
\label{fig:E-lz-sets}
\end{figure*}

In Fig.~\ref{fig:E-lz-sets} we show the evolution of the total specific energy and angular momentum components L$_z$ and L$_{perp}$ for individual GCs, obtained in TVP. As can be seen in upper row for 1--3 pre-sets, all of the cclusters mainly fall within the limits, especially in L$_z$ (\cite{Limberg2022, Malhan2022} and \cite{Sun2023}) boundaries. All of the clusters also mainly fall within the limits of the regions by energy from -17 to -8$\times$$10^4$ km$^2$ s$^{-2}$. In pre-set 4 the next five clusters do not correspond to the limits: NGC 6235, NGC 6426, and NGC 7078, and especially NGC 3201 and NGC 6101, which are completely out of the L$_z$ limits, $\approx$ -3$\times$$10^3$ kpc km s$^{-1}$.


\begin{table*}[ht]
\caption{Phase-space parameter limits for the GCs GE/S today, collected from the literature.}
\centering
\begin{tabular}{c|ccccccccc}
\hline
\hline
Source & L$_z$ & L$_{perp}$  & E & R$_{gal}$ & Z & r$_{\text{apo}}$ & r$_{\text{per}}$ & $a$ & {\tt ecc} \\ 
& $10^3$ kpc km s$^{-1}$ & $10^3$ kpc km s$^{-1}$ & $10^4$ km$^2$ s$^{-2}$ & kpc & kpc & kpc & kpc & kpc &  \\ 
\hline
\hline
\cite{Sun2023}      & [-0.8, 0.1]      & ---           & [-17, -8]      & --- & --- & ---        & --- & --- & >0.8 \\
\cite{Malhan2022}   &  [-0.715, 0.705]  & [0.08, 1.5]   & [-14.4, -11.6] & --- & --- & [16, 30]  & [1, 4] & [8.5, 17] & [0.7, 0.9]\\
\cite{Limberg2022}  &  [-0.5, 0.5]      & ---           & [-16, -10]     & --- & --- & ---        & --- & --- & --- \\
\cite{Massari2019}  &  [-0.8, 0.62]     & [0, 3.5]      & [-18.6, -9]    & --- & --- & < 25 & --- & --- & $\sim$ 0.8 \\ 
\hline
\hline
\end{tabular}
\label{tab:ges-GC-today}
\end{table*}

Analysing the bottom row in Fig.~\ref{fig:E-lz-sets}, it can be seen that all of the GCs in the first pre-set fall within the limits. In contrast to the evolution of the GCs in L$_z$, we can see clusters that do not correspond to the specific energy limits: NGC 7099 (black line) from the second pre-set, and NGC 5139, NGC 6205, and NGC 5897 (green, light blue, and beige) from the third pre-set. The GCs from the fourth pre-set -- NGC 6284, Terzan 10, NGC 4833, and NGC 3201, which falls within the boundaries by L$_{perp}$ -- are also not compatible with the specific energy limits. Most of them have, for almost all the evolutionary time, a larger bounding specific energy from -22 to -18.5$\times$$10^4$ km$^2$ s$^{-2}$. 

\begin{figure*}[htbp!]
\centering
\includegraphics[width=0.95\linewidth]{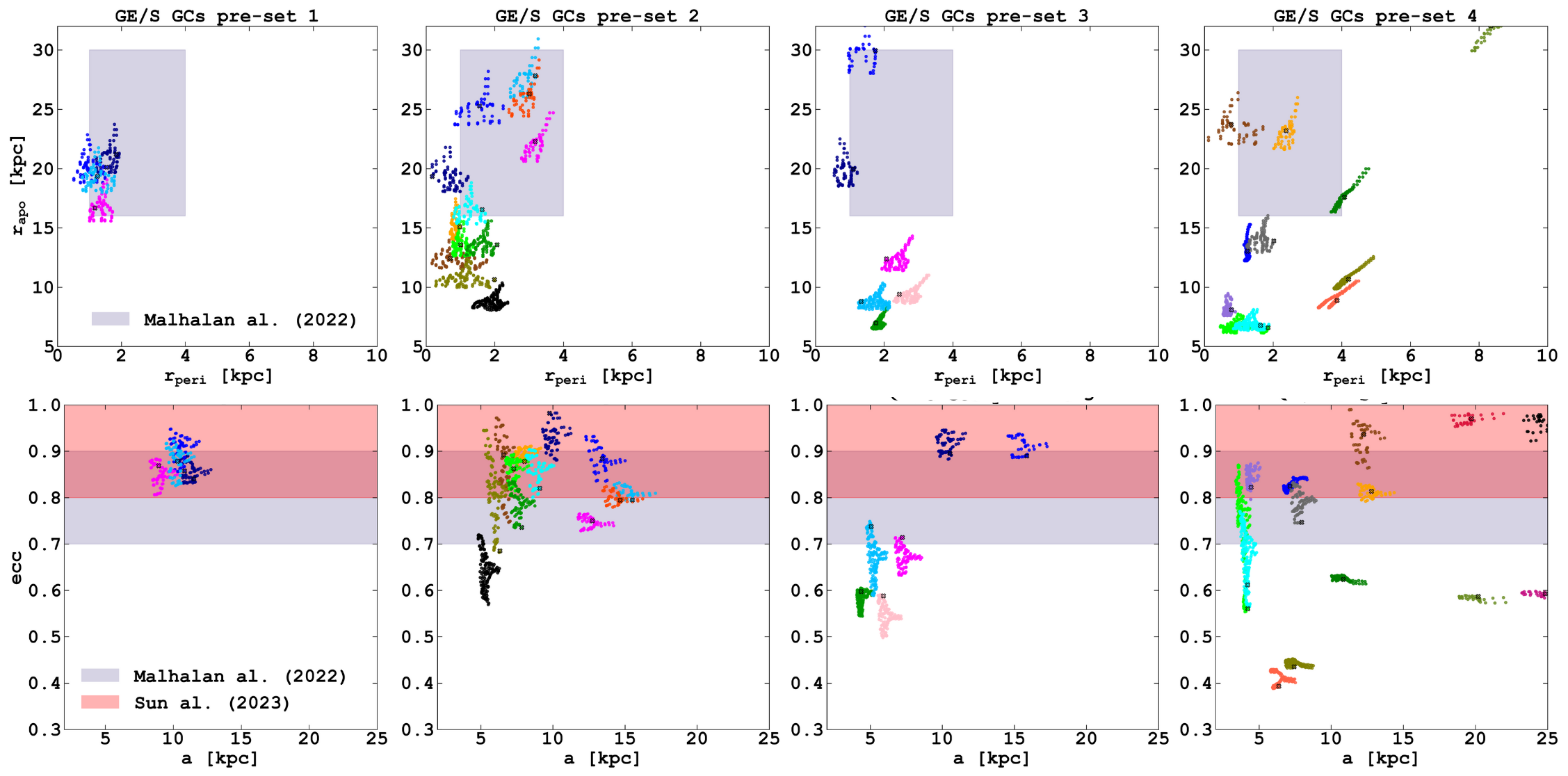}
\caption{Evolution of the GCs orbital parameters during the 9 Gyr lookback time integration in {\tt 411321} TVP external potential for four different pre-sets of GE/S GC's. The upper row shows the evolution of r$_{apo}$ and r$_{peri}$; the bottom row $a$ and {\tt ecc}. Transparent coloured areas represent the limit of orbital parameters for GE/S GCs based on the publications from Table~\ref{tab:ges-GC-today}. The GC colour coding corresponds to the legend from Fig.~\ref{fig:E-lz-sets}. Black dots show the GC position today. GCs Palomar 2, Palomar 15, and NGC 6101 are not shown in the bottom row for pre-set 4 due to their outside position on a plot boundaries (they have very large r$_{apo}$ values).}
\label{fig:a-e-sets}
\end{figure*}

Evolution of the orbital parameters during a 9 Gyr lookback time integration that we present in Fig.~\ref{fig:a-e-sets}. The coloured areas represent the distribution limits of the GC's GE/S for r$_{peri}$, r$_{apo}$ (upper row), together with $a$ and {\tt ecc} illustrated in the lower row. Note that the limits of the orbital elements are only defined only in two from our main four papers: see \cite{Malhan2022} and \cite{Sun2023}. 

For the pre-set 1, one can see a good correspondence of all the orbital parameters of individual GCs with the predefined ranges from the publications from Table~\ref{tab:ges-GC-today}. As can be seen for pre-sets 2--4, the clusters with {\tt ecc} < 0.7 completely out of the ranges. These are mainly: NGC 7099, NGC 6235, NGC 7078, NGC 6101, NGC 5139, NGC 5897, NGC 6205, NGC 288, and Terzan 10. The GC's distribution in the limits of r$_{peri}$ and r$_{apo}$ also show a large mismatch for almost all of the GCs in the second to fourth pre-sets. The only exceptions are a short list of GCs from the second pre-set that are almost entirely inside the r$_{peri}$ and r$_{apo}$ ranges: NGC 4147, 5634, 5904, 6864, and 7492.

\subsection{Resulting GE/S GCs sample}\label{subsec:res-sam}

Based on our numerical integration in the TVP external potential of the GCs, which we present in Figs.~\ref{fig:E-lz-sets} and ~\ref{fig:a-e-sets} with a corresponding analysis, we can summarised our pre-sets in the tree category. Namely, the first category represents the `most probable'. Here we represent the GCs that have a strong association with six parameters that determine the limits of GE/S: L$_{z}$, L$_{perp}$, the total energy, r$_{apo}$, r$_{peri}$, and {\tt ecc}. We get 15 GCs that fall into that category: NGC 1261, NGC 1851, NGC 7089, NGC 2808, NGC 2298, NGC 6864, NGC 6779, NGC 4147, NGC 5634, NGC 5904, NGC 7492, NGC 1904, IC 1257, NGC 6229, and NGC 6981. All of these clusters have a high conformity in orbital parameters and energy versus angular momentums during all 9 Gyr of integration of their orbital evolution. NGC 2808, NGC 1904, and IC 1251 do not satisfy the r$_{peri}$ limit according to \cite{Malhan2022}, but in our view it is not a critical parameter. This is first of all due to the quite large uncertainties in their definition and also the fact that this parameter was derived for our GC samples in only one publication \cite{Malhan2022}. The clusters we selected are also well suited in terms of chemical parameters. They are quite compatible with the GE/S general chemical pattern. Here only NGC 6229 and NGC 6981 can be under some discussion; see more details about the membership of these GCs to our `most probable' group in Sect. ~\ref{subsec:diss}.

We assume the second category to be `tentative'. Here, we include clusters that do not demonstrate full consistency in all six parameters, but that could correspond to the GE/S progenitor based on the total energy and angular momentum components, L$_{z}$, L$_{perp}$, or that have a significantly compatible chemical pattern with the GE/S system. In the end, we assume the next nine clusters to be `tentative': NGC 362, NGC 5286, NGC 6341, NGC 5897, NGC 7099, NGC 6426, ESO SC06, Djorg 1, and NGC 6426. The main reason for classifying the cluster in this group is its low energy value: < -18$\times$$10^4$ km$^2$ s$^{-2}$ (see Fig.~\ref{fig:E-lz-sets}, bottom row).  

Finally, in the `excluded' category we collected the clusters that in most of the parameters do not correspond to the specified limits or for which at least one of the parameters has a large discrepancy with the limits. In Table~\ref{tab:sam-res-gc} we present the resulting sample of GE/S clusters. In total, of the 36 GCs that were selected from the recent literature, we find 24 GCs that are the most probable and tentative candidates and that we are highly confident have a GE/S dwarf galaxy origin. Also in Table~\ref{tab:select-criter} we present detailed statistics on the orbital components and energy versus angular momentum values in phase-space in the context of the number of publications in which a particular cluster was mentioned.

In our current investigation, we use one of the five available MW-like TVP potentials selected from the IlustrisTNG-100 database; see Fig. 2 and Table 1 in \cite{Ishchenko2023a}. To make our results in Table \ref{tab:sam-res-gc} more robust, we selected an additional external potential -- {\tt 441327} -- and examined the impacts of these changes on the evolution of GCs' orbital parameters and energy versus momentum values. We carried out the same procedure as is described in Sect. \ref{sec:orb-reconstr} for our pre-selected GCs samples, collected in Table \ref{tab:gc-pre-sets}. In Figs. \ref{fig:E-lz-sets-441327} and \ref{fig:a-e-sets-441327}, we present the evolution of the GCs' phase-space parameters during the 9 Gyr lookback time integration in {\tt 441327} TVP external potential for our created pre-sets of GCs. 

We also provide a comparison of stellar mass loss in two simulations using both the time-dependent (TVP) and time-independent (TVP-FIX) potentials. We found slightly different mass loss, but during the whole simulation the difference between the TVP and TVP-FIX results is not more than a few percentage points; see the grey curves in Fig.~\ref{fig:t-relax-iniv-gs}. Also for more details see Appendix~\ref{sec:time-relax}.

Our additional tests incorporating a simplified dynamical friction model reveal that its influence on orbital evolution is modest over 9 Gyr, even for clusters with low pericentric distances. As is shown in Fig. \ref{fig:t-dyn}. The inclusion of dynamical friction causes only minor deviations in the galactocentric distance and orbital shape, without significantly altering energy or angular momentum; for more details see Appendix \ref{subsec:dyn-fr}.

By analysing the behaviour of orbital parameters, it is possible to identify clusters that demonstrate an even more striking discrepancy with the boundaries of the GE/S; namely, NGC 5139, NGC 6205, NGC 5897 from third pre-set and NGC 4833, NGC 6284, Terzan 10, NGC 6584, and NGC 7078, which have an even stronger bound connection with the Galaxy core than in the {\tt 411321} TVP. However, aside from the stronger bounding energies, the orbital and angular momentum parameters do not change significantly in the {\tt 441327} potential relative to the {\tt 411321} potential. So, the distribution of the orbital parameters and angular-momentum variables of the `most probable' and `tentative' clusters satisfies all the limits of Tab. ~\ref{tab:ges-GC-today}.

\begin{table}[htbp]
\caption{Compiled sample of GE/S GC's based on Figs.~\ref{fig:a-e-sets},~\ref{fig:E-lz-sets} and~\ref{fig:3d-phase-space} with Table~\ref{tab:select-criter}.}
\centering
\sisetup{separate-uncertainty}
\resizebox{0.50\textwidth}{!}{
\begin{tabular}{c|c}
\hline
\hline
Category  &  GC's Name \\
\hline
\hline
\raisebox{-4.5ex}[0pt][0pt]{\centering Most probable}
  &  NGC 1261, NGC 1851, NGC 2298, NGC 7089, \\
  &  NGC 1904, NGC 2808, NGC 4147, NGC 5634, \\
  &  NGC 5904, NGC 7492, NGC 6229, NGC 6864, \\
  &  IC 1257, NGC 6981, NGC 6779, \\
\hline
\raisebox{-3ex}[0pt][0pt]{\centering Tentative}
  &  NGC 362, NGC 5286, NGC 6341, NGC 6584,   \\
  &  NGC 6426, NGC 7099, ESO SC06, Djorg 1,    \\
  &  NGC 5897    \\

\hline
\raisebox{-3ex}[0pt][0pt]{\centering Excluded}
  &  NGC 288, NGC 5139, NGC 6205, NGC 3201,    \\
  &  NGC 4833, NGC 6101, NGC 6235, NGC 6284,    \\
  &  NGC 7078, Terzan 10, Palomar 2, Palomar 15    \\
\hline
\end{tabular}}
\label{tab:sam-res-gc}
\end{table} 

\begin{figure}[htbp!]
\centering
\includegraphics[width=0.85\linewidth]{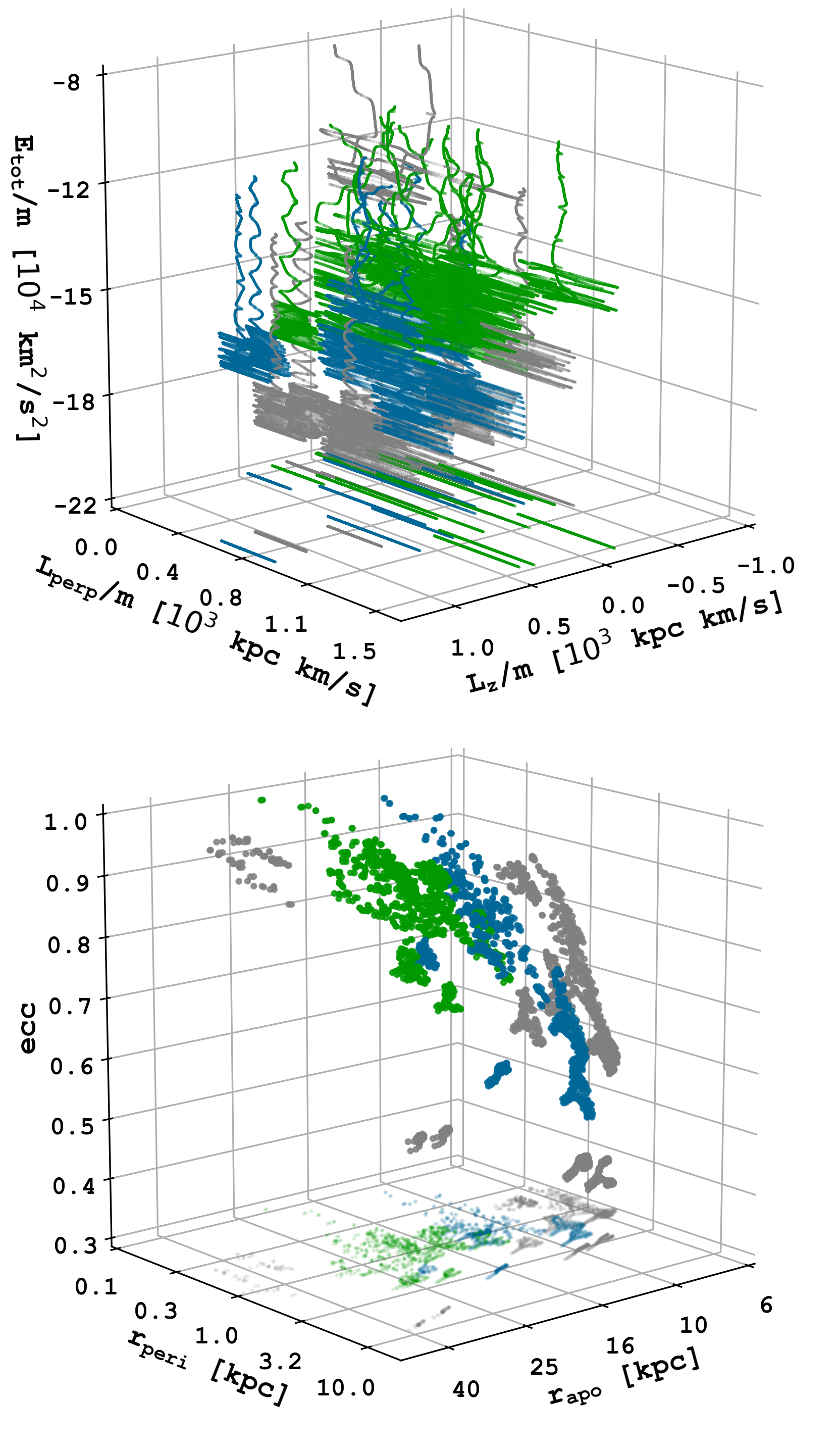}
\caption{Evolution of the GC's energy angular momentum and orbital parameters, during the 9 Gyr lookback time integration in the {\tt 411321} TVP external potential. The green colour represents the most probable GCs of the GE/S category, the blue are tentative, and the grey excluded GCs.}
\label{fig:3d-phase-space}
\end{figure}

We also present summaries of the 3D plots in Fig.~\ref{fig:3d-phase-space}, with the GCs distribution in the context of phase-space parameters: L$_{z}$, L$_{perp}$, and the total energy and orbital parameters, r$_{apo}$, r$_{peri}$, and {\tt ecc}. For each GC we represent the time evolution during the integration up to 9 Gyr. We clearly see the separation between our tree categories, especially in orbital parameters. The GCs from the `most probable' and `tentative' categories have, as was expected, high values in the {\tt ecc}: 0.7--0.99, while the `excluded' category demonstrate the large variation in {\tt ecc} from 0.3 to 0.9. Also GCs from this category appear to have a low total energy: from -22 to -18$\times$$10^4$ km$^2$ s$^{-2}$. The other two categories range from -18 to -12$\times$$10^4$ km$^2$ s$^{-2}$ during their entire evolution, which is closer to the GE/S stream distribution in the Galaxy.

\subsection{GCs under discussion}\label{subsec:diss}

Here, we would like to discuss in more detail our resulting lists in Table \ref{tab:sam-res-gc}. The clusters NGC 6229 and NGC 6981, which we assume to be the `most probable', stand out from the other GCs in the group due to the lack of their chemical abundance observations; see Table ~\ref{tab:select-criter}. Note that the authors of references \cite{Massari2019} and \cite{Malhan2022} consider NGC 6229 and NGC 6981 to correspond to the GE/S progenitor, whereas a later publication (\cite{Callingham2022}) associates these GCs with the group of the Helmi stream. On the other hand, they demonstrate a strong association with GE/S in all kinematic parameters also based on our own numerical simulations in TVP.

In the tentative category we want to note a group of clusters -- NGC 362, NGC 5286, NGC 5897, NGC 6341, and NGC 7099 -- that have a strong chemistry association with GE/S, but kinematic parameters that have lower energy values (from -20 to -16$\times$$10^4$ km$^2$ s$^{-2}$) compared with others GCs here. Almost all of the objects from this group have quite similarly shaped orbits. For example, NGC 5286 and NGC 5897 illustrate well the orbits from this group (see, Fig.~\ref{fig:orb-plots}). For these GCs we have a typically strong and multiple passages in the Galactic thick disc. During the GE/S accretion event, it is possible that some clusters, before they become a part of our Galaxy, were stripped by some dwarf galaxies (which themselves earlier were accreted by the MW) and as a result they can form a subgroup of GCs inside the current GE/S stream. These objects, as a result of such a subsequent mergers, can have stronger bounding energies, smaller apocentre distances, and also lower eccentricities.

Note that the authors in \cite{Malhan2022} classified this group (the five systems mentioned above) as members of the Pontus dwarf galaxy. However, the subsequent analysis in \cite{CARMA2025b} shows that this group is indeed strongly associated with GE/S by age and also metal abundances. 

GC Djorgovsky 1 has much in common with GE/S clusters in kinematic parameters. However, 
we find only one publication that has studied the chemical abundances in this object; see Table \ref{tab:select-criter}. So we consider to assume this cluster to be in the tentative group of our sample.

In the excluded group, we also have a few objects with some specific characteristics. With regard to the clusters NGC 288 and NGC 6205 shown in \cite{CARMA2025a} [Fig. 3], the authors found that these two clusters were much older than the GE/S dwarf galaxy. At the same time, the kinematic coincidence of the parameters of the clusters with GE/S can be explained by the fact that during the falling of the GE/S galaxy, the orbits of these two clusters were significantly perturbed \citep{CARMA2025b}. Based on these speculations, we excluded them from being probable GE/S stream members.

We also need to explain, in more detail, why in the excluded group we have Palomar 2 and Palomar 15. These clusters have angular momentum and energy values that are very typical of GE/S clusters. Also we have a quite limited knowledge of the chemical composition of Palomar 2 and 15 (similar to the case of Djorgovsky 1). We exclude these objects from being probable GE/S members, mainly based on their quite large semi-major distances ($\sim$20 - 25 kpc) and high eccentricity values ($\sim$0.95); see the bottom row of Fig.~\ref{fig:a-e-sets}. 

NGC 5139 (also known as Omega Centauri) is the most discussed cluster. Some of the authors \citep{Massari2019, Kruijssen2020} consider this object to be a possible stellar core of the former GE/S dwarf galaxy or Sequoia dwarf based on the age-metallicity relation. In Fig.~\ref{fig:orb-plots} we also present the orbit of this cluster. In a sense of binding energy, this object is one of the most bound systems in our sample (Fig.~\ref{fig:E-lz-sets}). At the same time, as can be seen from the Table~\ref{tab:select-criter}, this cluster has typical angular momentums compatible with GE/S only in L$_Z$. In the absence of other compelling factors, we consider leaving this cluster in the excluded group.

We also note that in the most probable sample we have ten GCs that have clockwise rotation and five that have anti-clockwise rotation (in the Cartesian Galactocentric reference frame), while in the tentative sample six GCs demonstrate anti-clockwise rotation. At the same time, the author states that the GE/S dwarf galaxy was falling towards our Galaxy in an orbit with anti-clockwise rotation \cite{Helmi2018, Koppelman2020}.

\section{N-body modelling of the GCs GE/S members} \label{sec:n-body-model}

In this section we carry out numerical modelling of 24 GCs in order to estimate the distribution of stars in the phase-space that no longer belong to the GE/S clusters today.

\subsection{GCs GE/S initial conditions and integration procedure}\label{sec:init-gc}

The aim of this $N$-body simulation is to study the distribution of stars in the Galaxy that have been lost by the GCs GE/Ss due to their own relaxation processes during the full 9 Gyr orbital evolution. Since our primary focus in the modelling was the dynamical behaviour of escaped, sub-solar-mass, long-lived stars within the cosmological timescale and a varying external Galactic gravitational potential (IllustrisTNG-100 motivated -- {\tt 411321} TVP). In the current simulation, we turned off the individual particles' stellar evolution during the $N$-body simulations. While a detailed investigation into GC's internal stellar mass loss evolution would indeed require the incorporation of the stellar evolution module, our current study specifically targets stars with lifetimes of the order of 9 Gyr and more. These objects can travel, after escape from GCs, independently within the whole Galaxy. For this specific subset of stars, the time-dependent stellar evolution mass loss can be reasonably neglected. A similar approach was already successfully applied in our previous paper \cite{Ber2024} in which we study the dynamical evolution of the escaped stars from the Reticulum II dwarf galaxy. 

As a basic initial physical model for all the 24 GCs at 9 Gyr ago, we chose the equilibrium King model \citep{King1962} with the concentration parameter W$_0$ = 8.0 and with the fixed total mass of $10^5$ M$_{\rm \odot}$. We chose these parameters to initially have a maximally compact King object in terms of the half-mass radius to tidal radius ratio. This allows us to have systems with the maximum lifetime and long survivability in the TVP external tidal field. 

We varied the clusters' remaining two physical parameters; namely, the number of particles and the initial half-mass radius, depending on the types of cluster orbits inside the Galaxy. Thus, for clusters whose orbits are relatively close to the centre of the Galaxy (r$_{apo}$ less then 10 kpc) we used a half-mass radius of 2 pc and the initial number of particles N = 45--50\textit{k} (depending on the clusters current masses). 

For clusters that have long, elongated radial orbits and a small number of passages near the Galactic centre, we applied models with a larger half-mass radius of 4 pc and with the initial number of particles 40--45\textit{k}. Using these two physical parameters (half-mass radius and initial number of particles), we tried to provide the more or a less comparable stellar escape rate of low-mass stars from the clusters with significantly different initial masses. 

Besides low particle numbers in our $N$-body models (compared to the real number of stars in the Galactic GCs), we managed to provide the long-term (up to 9 Gyr) survivability of our dynamical models, which allow us to have a long term `source' of escaped stars from these clusters. We used these stars in our study as `indicators' (tracers) of the long-term orbital motion of our selected GCs. This approach -- using a relatively low $N$ for each GCs -- enables us to run a significantly larger set of $N$-body simulations within a realistic computer running time frame.

To evaluate the possible influence of initial conditions on the dynamical evolution of GCs, we performed a number of additional simulations with different numbers of particles for several GCs from our list. We discuss the evolution of the star loss due to the orbital motion and relaxation time for these GCs in Appendix~\ref{sec:time-relax}.

\subsection{Distribution of stars in phase-space at the present time from the GE/S GCs sample}\label{sec:gc-distr-ps}
  
\begin{figure*}[]
\centering
\includegraphics[width=0.99\linewidth]{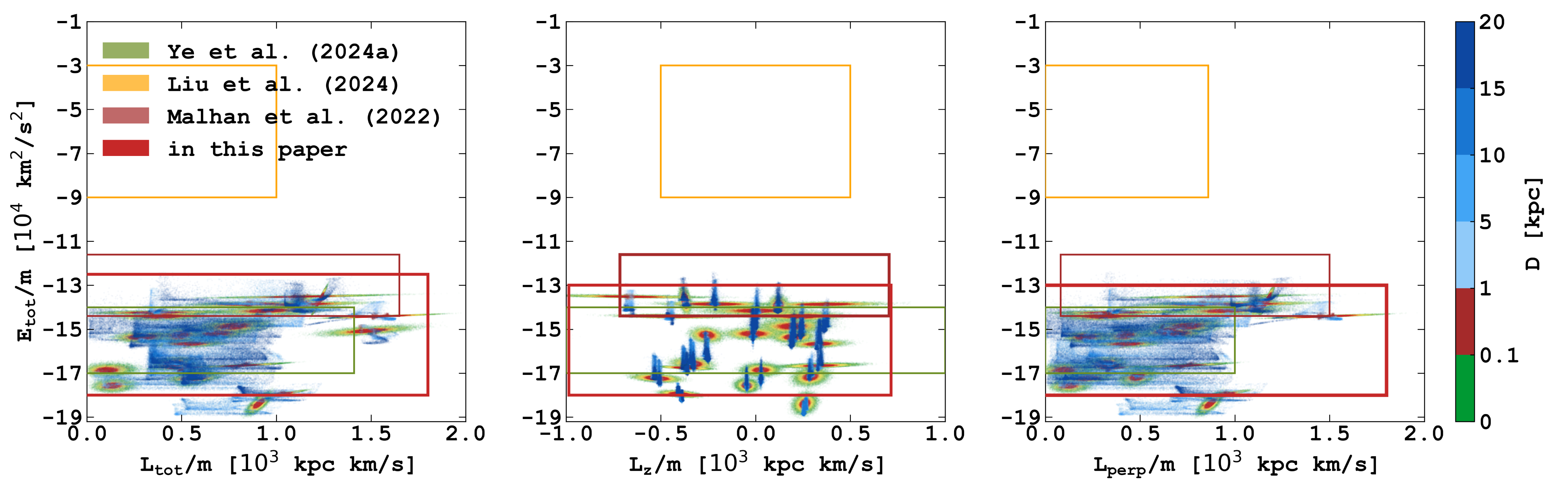}
\caption{GE/S GC stars distribution in phase-space for today for two samples (combined most probable with tentative 
 from Table~\ref{tab:sam-res-gc}). Patches represent the GE/S stream phase-space limits from Table~\ref{tab:ges-stream-today}. The colour-coding represents the relative distance from the GC density centre.}
\label{fig:phase-today+stream}
\end{figure*}

In Fig.~\ref{fig:phase-today+stream} we show the distribution of stars in the Galaxy at the present time, based on our sample of 24 GCs from the GE/S groups (most probable and tentative) during the 9 Gyr of their evolution. As we see, the global stellar distribution (escaped stars and stars that still belong to the cluster) in the L$_{\rm tot}$ space have quite a wide distribution range: from 0 to 1.8$\times$10$^3$ kpc km$^1$ s$^{-1}$ and from -18 to -12.2$\times$10$^4$ km$^2$ s$^{-2}$ in energy values. In the L$_{\rm z}$ component, as was expected, we see a quite narrow distribution of all stars for each GC. Most of these stars today are already not bound gravitationally to the progenitor GC itself. Precisely these unbound stars form our GE/S stellar debris in the phase-space region; see Fig.~\ref{fig:phase-today+stream} (thick red box) and `in this paper' in Table~\ref{tab:ges-stream-today}. 

As we see from Fig.~\ref{fig:phase-today+stream},  the densest stellar regions overlap quite well with the possible limits of the GE/S stream in the energy--angular momentum phase-space, presented in the publications of \cite{Ye2024} and \cite{Malhan2022}. As a side remark, we need to note that the limits used in the paper \cite{Liu2024} are obviously not compatible with our GCs GE/S stellar debris distribution.

\begin{table*}[]
\caption{Phase-space parameters for the GE/S stream at today.}
\centering
\begin{tabular}{c|cccccccc}
\hline
\hline
Source & L$_z$ & L$_{perp}$  & E & R$_{gal}$ & |Z| & r$_{\text{apo}}$ & r$_{\text{per}}$ & {\tt ecc} \\ 
& $10^3$ kpc km s$^{-1}$ & $10^3$ kpc km s$^{-1}$ & $10^4$ km$^2$ s$^{-2}$ & kpc & kpc & kpc & kpc &  \\ 
\hline
\hline
\cite{Ye2024}  &  [-1; 1] & [0; 1] & [-18; - 14] & < 30 & [3; 14] &[10; 30] & [0; 4] & [0.7; 1.0] \\
\cite{Liu2024}   & [-0.5; 0.5] & [0; 0.86] & [-9; - 3] & [3; 15] & --- & --- & --- & $\sim1$ \\  
\cite{Malhan2022}    &  [-0.715; 0.705] & [0.08; 1.5]  & [-14.4; -11.6] & --- & --- &  [16; 30] & [1; 4] & [0.7; 0.9]\\
\hline
In this paper &  [-0.98; 0.72] & [0; 1.8]  & [-18; -12.2] & <18 & <15 & [10; 28] & [1; 4] & >0.7 \\
\hline
\end{tabular}
\label{tab:ges-stream-today}
\end{table*}

\section{Phase-space comparison of the GE/S GCs' escaped stars with the metallicity-selected  individual GE/S stars}\label{sec:compar-vpm}

Our comparison is based on the GE/S VMP star sample, which are investigated in detail in recent publications: \cite{Ye2024}, \cite{Ernandes2024}, and \cite{Molaro2020}. All of these samples focus on their selection criteria of stars that are connected with the GE/S galaxy based on chemical abundances. We make a comparison between these GE/S stellar samples (selected by chemistry) and our resulting energy and angular momentum distribution of the stars from GCs of GE/S for today in our dynamical external TVP. 

\begin{figure*}[bp!]
\centering
\includegraphics[width=0.90\linewidth]{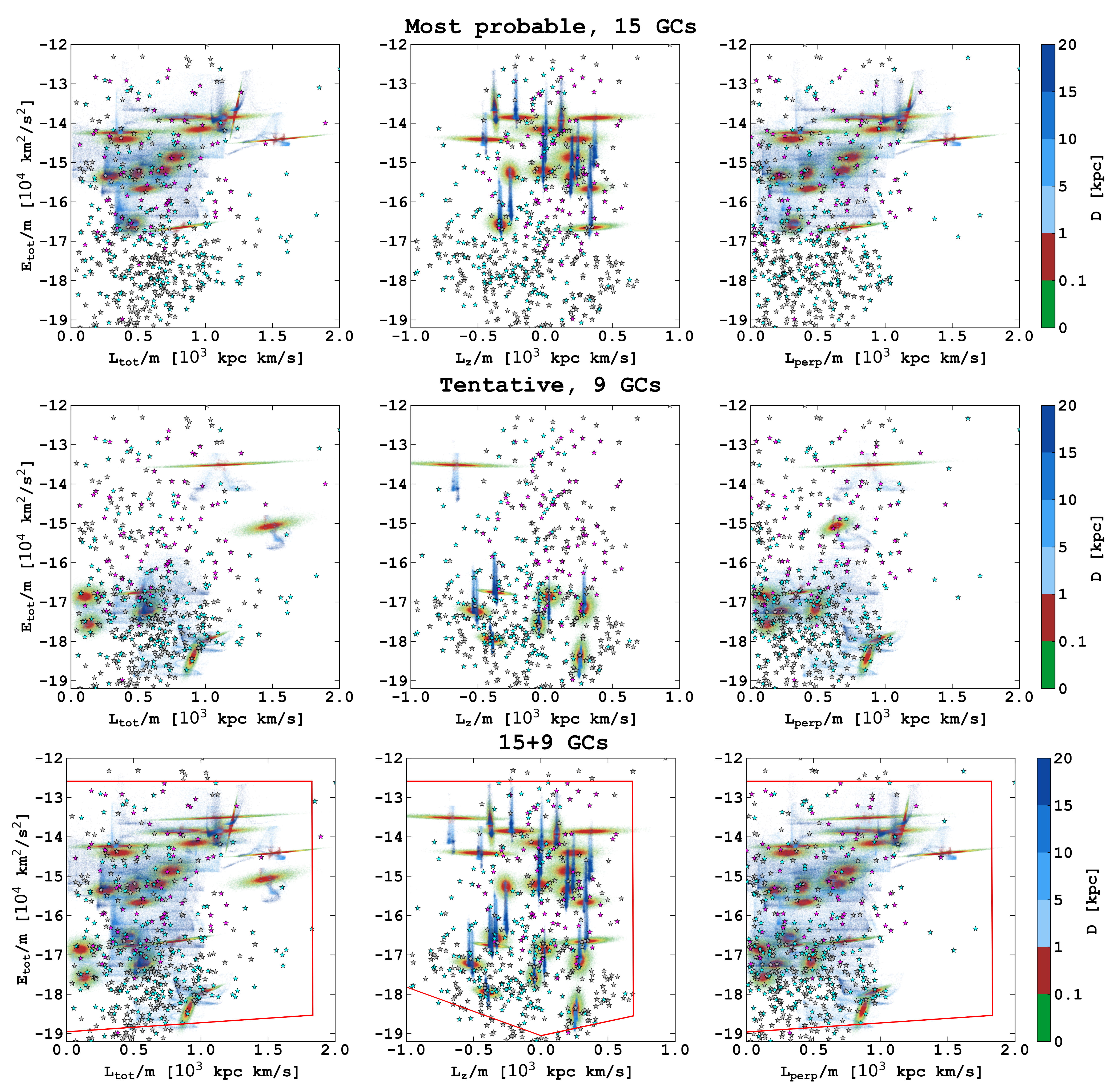}
\caption{Distribution of the ejected stars from the GE/S GCs together with the selected VMP stars. The colour-coding bar for GCs represents the relative distance from the GC density centre itself. Cyan star symbols represent GE/S stars from \cite{Molaro2020}, magenta -- \cite{Ernandes2024}, and grey -- the \cite{Ye2024} sample, respectively.}
\label{fig:den-todat}
\end{figure*}

\paragraph{GE/S stars from \cite{Ye2024}.} This study (besides the chemical selection criteria) is based on the VMP stellar samples using kinematic and action-angle analysis of the individual stars' orbits. The authors analysed the VMP stars sample based on the \textit{Gaia} DR3 catalogue, together with the application of the radial velocities from the LAMOST DR9 catalogue. In summary, the authors found 1070 stars that are probably associated with the original GE/S progenitor. From this sample we obtained 343 stars with full 6D astrometry data, available in the \textit{Gaia} DR3 catalogue. Another 727 GE/S VMP stars have low-accuracy (compared to \textit{Gaia} DR3) radial velocities based on the LAMOST catalogue.

\paragraph{GE/S stars from \cite{Ernandes2024}.} Here, the GE/S VMP stellar sample based on the data from the SAGA base. The authors make a stellar sample based on the metallicity abundance range that is expected for GE/S stars: as -2.2 < [Fe/H] < -0.5. From this sample, we obtained 73 stars, which have \textit{Gaia} DR3 astrometry data. 

\paragraph{GE/S stars from \cite{Molaro2020}.} The authors investigated the stellar abundances {\tt Li} and {\tt Be} from the possible members of GE/S. In this work, to identify stars that are connected with GE/S progenitor, the authors additionally defined the limits in the angular momentum L$_z$ component. Note that the sample is based on the data from \textit{Gaia} DR2 catalogue together with the APOGEE sky survey. From this sample we collected 171 stars, with \textit{Gaia} DR3 astrometry data. 

The total distribution of the combined 587 GE/S stars presented in Fig.~\ref{fig:den-todat} in energy and angular momentum co-ordinates. To calculate these values for each star, we used their 6D kinematic data from \textit{Gaia} DR3, and our external Galactic potential {\tt 411321} TVP. In Fig.~\ref{fig:den-todat} we plot these 587 GE/S stars together with the ejected stars distribution of GE/S GCs for today. The \cite{Ernandes2024} sample shows the greatest overlap with the phase-space distribution of the ejected stars from our two GE/S GCs categories. In the top row we put the 15 most probable GE/S GCs. In the middle row we put the nine tentative GE/S GCs. In the bottom row, we show the combination of these two GCs samples in one figure. The red line on the bottom panels represents the limits in the energy and angular momentum space, according to our modelled ejected stars from the GCs of the GE/S. The corresponding values of the possible GE/S debris boundaries we also present in the Table \ref{tab:ges-stream-today}. As one can see from Fig.~\ref{fig:den-todat}, all the GE/S stellar samples overlap quite well with the groups of escaped stars from the GE/S GCs. 

\section{Conclusions}\label{sec:discus}

High-precision astrometry data from \textit{Gaia} in the last few years has allowed us to identify the most massive stellar debris that was supposed to be the remnant of the major merger with the dwarf galaxy named GE/S \citep{Helmi2018}. Further investigations allowed us to combine stellar streams from such events with GCs. Based on the recent literature review, we collected and combined catalogue of the GCs, which potentially can have the GE/S dwarf galaxy origin. We performed $N$-body backward integrations for 36 collected from literature GCs, up to the probable time of the merging event of the GE/S with our MW, i.e. up to 9 Gyr. We carried out the GC orbital reconstruction in the time-variable external MW-like galactic potentials, selected from the IlustrisTNG-100 database. Based on Figs. \ref{fig:E-lz-sets} and \ref{fig:a-e-sets}, we selected GE/S GCs using the limits of their distribution in orbital parameters and energy with angular momentum for GE/S GCs that we find in the literature; see Table \ref{tab:ges-GC-today}. 

Note that applying a time-varying potential to define the boundaries of GCs offers a more flexible and realistic framework compared to static potentials. Unlike static models, which constrain GCs to fixed energy and angular momentum values and thus impose stricter membership criteria, for the case with time-dependent potential we have more natural variations in orbital parameters and as a result more variability in the membership criteria too. 

Therefore, we analysed the evolution of orbital parameters and angular momentum and created a sample of 24 GCs that most probably have an association with the GE/S progenitor. We distributed the collected GCs into two categories: most probable and tentative. As a result we have a total list of 24 individual GCs: NGC 1261, NGC 1851, NGC 2298, NGC 7089, NGC 1904, NGC 2808, NGC 4147, NGC 5634, NGC 5904, NGC 7492, NGC 6229, NGC 6864, IC 1257, NGC 6981, NGC 6779, NGC 362, NGC 5286, NGC 6341, NGC 6584, NGC 6426, NGC 7099, ESO SC06, Djorg 1, and NGC 5897; see Table \ref{tab:sam-res-gc}. In the next step, we ran a set of dynamical $N$-body simulations to study the stellar ejection processes due to their dynamical evolution and orbital motion in the TVP external potential.

We identified a possible dynamical connection between individual stars with extremely low metallicities (VMP stellar samples) and stars, ejected from GCs that exhibit a significant connection with GE/S; see Fig. \ref{fig:den-todat}. This connection is explored using the current phase-space information for these stars, which provides detailed insights into their positions and velocities within the MW, applying the time-variation external Galactic potential. Furthermore, we applied the advanced dynamical star loss model to simulate the behaviour of GCs together with the ejected stars that were associated with the GE/S over a period of 9 Gyr of orbital motion. This model allows us to understand the evolution of the our 24 selected GCs, including tidal stripping and other gravitational influences, which affect the global stellar distribution over time.

We defined the GE/S debris limits based on the escaped stars from our selected 24 GE/S GCs that were dynamically modelled in external time-variable Galactic potential. Based on this modelling, we obtain the next phase-space GE/S limits: specific energy -18 to -12.2 $\times10^4$ km$^2$ s$^{-2}$, in L$_{\rm z}$ from -0.98 to 0.72 $\times10^3$ kpc km s$^{-1}$; for L$_{\rm perp}$: from 0 to 1.8 $\times10^3$ kpc km s$^{-1}$; see Fig.~\ref{fig:den-todat}. Also we redefined the possible limits of the GE/S GCs in Galactic area with the following orbital parameters: an apocentre distance of 10--28 kpc, a pericentre distance of 1--4 kpc, and limits of < 18 kpc in the Galactocentric radius and < |15| kpc in the $Z$ direction; see Table~\ref{tab:ges-stream-today}. 

Our current study shows us a possible future extension of our models with the inclusion of the more recent IllustrisTNG-50 cosmological database for the detailed simulation of proto-MW and GE/S mergers. To conduct a detailed study on the origin of GCs, it is also necessary for us to create a comprehensive model of the two galaxies (proto-MW and GE/S), including their GCs subsystems too. This complex model of the GE/S merger involves the highest possible particle resolution, with GCs acting as reference points to track the flow of the merger. In particular, we are interested in the distribution of GCs in phase space 9–11 Gyr ago. This will enable us to create a more robust sample of GE/S GCs. At the same time, it will address the question of whether GCs in phase-space can provide solid evidence of the host galaxy's accretion history.

\begin{acknowledgements}

The authors thank the anonymous referee for a very constructive report and suggestions that helped significantly improve the quality of the manuscript.
MI and PB thanks the support from the special program of the Polish Academy of Sciences and the U.S. National Academy of Sciences under the Long-term program to support Ukrainian research teams grant No.~PAN.BFB.S.BWZ.329.022.2023. The work was carried out with the support of the Ministry of Science and Higher Education of the Republic of Kazakhstan within the framework of Projects No AP26100669 ``Galactic Archaeology of Gaia-Enceladus Globular Clusters as Indicators of the Formation History of the Milky Way Disk''.

\end{acknowledgements}

\bibliographystyle{mnras}  
\bibliography{gc-ge}   

\begin{appendix}

\onecolumn

\section{Estimation of the GC measurement errors}\label{sec:gaia-errors}

Here we in detail investigated the error measurements and their influence on the orbital evolution of GCs in the past. In Table ~\ref{tab:cur-pos} we present the 6D kinematic parameters for selected GCs and their computed relative errors that we take from the catalogue \cite{Baumgardt2021} and on the web page\footnote{\url{https://people.smp.uq.edu.au/HolgerBaumgardt/globular/orbits.html}}. All astrometry data is based on \textit{Gaia DR3}. It should be noted that the observational data (R$_\odot$, R$_V$, $\mu_\alpha$, and $\mu_\delta$) have corresponding measurement errors, this is especially true for radial velocity and parallax of the GC's. Such errors can lead to large uncertainties during reconstruction of the GC orbits on a cosmological timescale. For some of the GCs, the orbit may have large discrepancies in its shape. As can be seen, most GCs have a high precision in their measured kinematic parameters. In general, the relative errors are below 2\%. However, we also see some examples for larger errors. For example, NGC 6584 has in $\mu_\alpha$ a relative error of 10\%. The NGC 7089 has an error in the radial velocity R$_V$ that is 7. 9\%.

\begin{table*}[h]
\setlength{\tabcolsep}{4pt}
\centering
\caption{6D kinematic properties at the current time for GCs.}
\label{tab:cur-pos}
\resizebox{1.00\textwidth}{!}{
\begin{tabular}{ccccccccccc}
\hline
\hline 
 Name        & RA    & Dec        & R$_\odot$ $\pm$ eR$_\odot$ & $\Delta $R$_\odot$ &  R$_V$ $\pm$ eR$_V$ &  $\Delta$R$_V$ & $\mu_\alpha$ $\pm$ e$\mu_\alpha $ &   $\Delta \mu_\alpha $ &   $\mu_\delta$ $\pm$ e$\mu_\delta$ &   $\Delta\mu_\delta$ \\
  &  deg & deg & kpc  &    \% &  km\ s$^{-1}$  &   \% &  mas/yr &        \% & mas/yr &        \% \\
(1) & (2) & (3) & (4) & (5) & (6) & (7) & (8) & (9) & (10) & (11) \\
\hline
\hline
 NGC\ 1261     &  48.0675 & -55.2162  & $16.40 \pm 0.19$ &      1.1 & $71.34 \pm 0.21$   &      0.3 & $1.60 \pm 0.01$  &        0.6 & $-2.06 \pm 0.01$ &        0.44 \\
 NGC\ 1851     &  78.5282 & -40.0466  & $11.95 \pm 0.13$ &      1.1 & $321.40 \pm 1.55$  &      0.5 & $2.14 \pm 0.01$  &        0.4 & $-0.67 \pm 0.01$ &        1.19 \\
 NGC\ 2298     & 102.248  & -36.0053  & $9.83 \pm 0.17$  &      1.7 & $147.15 \pm 0.57$  &      0.4 & $3.31 \pm 0.01$  &        0.2 & $-2.18 \pm 0.01$ &        0.32 \\
 NGC\ 7089     & 323.363  &  -0.82325 & $11.69 \pm 0.11$ &      0.9 & $-3.78 \pm 0.30$   &      7.9 & $3.44 \pm 0.01$  &        0.3 & $-2.18 \pm 0.01$ &        0.41 \\

\hline
 NGC\ 362      &  15.8094 & -70.8488  & $8.83 \pm 0.10$  &      1.1 & $223.12 \pm 0.28$  &      0.1 & $6.69 \pm 0.01$  &        0.1 & $-2.53 \pm 0.01$ &        0.36 \\
 NGC\ 1904     &  81.0458 & -24.5244  & $13.08 \pm 0.18$ &      1.4 & $205.76 \pm 0.20$  &      0.1  & $2.47 \pm 0.01$  &        0.3 & $-1.60 \pm 0.01$ &        0.5  \\
 NGC\ 2808     & 138.013  & -64.8635  & $10.06 \pm 0.11$ &      1.1 & $103.57 \pm 0.27$  &      0.3 & $1.00 \pm 0.01$  &        0.7  & $0.27 \pm 0.01$  &        2.6  \\
 NGC\ 4147     & 182.526  &  18.5426  & $18.54 \pm 0.21$ &      1.1 & $179.35 \pm 0.31$  &      0.2 & $-1.69 \pm 0.01$ &        0.8 & $-2.09 \pm 0.01$ &        0.62 \\
 NGC\ 5286     & 206.612  & -51.3742  & $11.10 \pm 0.14$ &      1.3 & $62.38 \pm 0.40$   &      0.6 & $0.19 \pm 0.01$  &        3.7 & $-0.16 \pm 0.01$ &        4.38 \\
 NGC\ 5634     & 217.405  &  -5.97643 & $25.96 \pm 0.62$ &      2.4 & $-16.07 \pm 0.60$  &      3.7 & $-1.69 \pm 0.01$ &        0.5 & $-1.49 \pm 0.01$ &        0.54 \\
 NGC\ 5904     & 229.638  &   2.08103 & $7.48 \pm 0.06$  &      0.8  & $53.50 \pm 0.25$   &      0.5 & $4.07 \pm 0.01$  &        0.2 & $-9.87 \pm 0.01$ &        0.06 \\
 NGC\ 6341     & 259.281  &  43.1359  & $8.50 \pm 0.07$  &      0.8 & $-120.55 \pm 0.27$ &      0.2 & $-4.93 \pm 0.01$ &        0.2 & $-0.63 \pm 0.01$ &        1.9  \\
 NGC\ 6779     & 289.148  &  30.1835  & $10.43 \pm 0.14$ &      1.3 & $-136.97 \pm 0.45$ &      0.3 & $-2.00 \pm 0.01$ &        0.4 & $1.62 \pm 0.01$  &        0.43 \\
 NGC\ 6864     & 301.52   & -21.9212  & $20.52 \pm 0.45$ &      2.2 & $-189.08 \pm 1.12$ &      0.6 & $-0.61 \pm 0.02$ &        2.6 & $-2.80 \pm 0.01$ &        0.54 \\
 NGC\ 7099     & 325.092  & -23.1799  & $8.46 \pm 0.09$  &      1.1 & $-185.19 \pm 0.17$ &      0.1 & $-0.73 \pm 0.01$ &        1.1 & $-7.31 \pm 0.01$ &        0.1  \\
 NGC\ 7492     & 347.111  & -15.6115  & $24.39 \pm 0.57$ &      2.3 & $-176.70 \pm 0.27$ &      0.1 & $0.78 \pm 0.01$  &        1.4 & $-2.32 \pm 0.01$ &        0.47 \\
\hline
 NGC\ 288      &  13.1885 & -26.5826  & $8.99 \pm 0.09$  &      1.0    & $-44.45 \pm 0.13$  &      0.3 & $4.16 \pm 0.00$  &        0.1  & $-5.71 \pm 0.00$ &     0.07 \\
 NGC\ 5139     & 201.697  & -47.4795  & $5.43 \pm 0.05$  &      0.9 & $232.78 \pm 0.21$  &      0.1 & $-3.24 \pm 0.01$ &        0.3 & $-6.73 \pm 0.01$ &        0.16 \\
 NGC\ 5897     & 229.352  & -21.0101  & $12.55 \pm 0.24$ &      1.9 & $101.31 \pm 0.22$  &      0.2 & $-5.42 \pm 0.01$ &        0.1 & $-3.39 \pm 0.01$ &        0.18 \\
 NGC\ 6205     & 250.422  &  36.4599  & $7.42 \pm 0.08$  &      1.1 & $-244.90 \pm 0.30$ &      0.1 & $-3.14 \pm 0.01$ &        0.2 & $-2.57 \pm 0.01$ &        0.23 \\
 NGC\ 6229     & 251.745  &  47.5278  & $30.11 \pm 0.47$ &      1.5 & $-137.89 \pm 0.71$ &      0.5 & $-1.16 \pm 0.02$ &        1.5 & $-0.46 \pm 0.02$ &        3.9  \\
 IC\ 1257      & 261.785  &  -7.09306 & $26.59 \pm 1.43$ &      5.4 & $-137.97 \pm 2.04$ &      1.5 & $-1.03 \pm 0.03$ &        2.4 & $-1.49 \pm 0.02$ &        1.28 \\
\hline
 NGC\ 3201     & 154.403  & -46.4125  & $4.74 \pm 0.04$  &      0.8 & $495.38 \pm 0.06$  &      0.01 & $8.35 \pm 0.01$  &        0.1 & $-1.98 \pm 0.01$ &        0.25 \\
 NGC\ 4833     & 194.891  & -70.8765  & $6.48 \pm 0.08$  &      1.2 & $201.99 \pm 0.40$  &      0.2  & $-8.39 \pm 0.01$ &        0.1 & $-0.97 \pm 0.01$ &        0.62 \\
 NGC\ 6101     & 246.45   & -72.2022  & $14.45 \pm 0.19$ &      1.3 & $366.33 \pm 0.32$  &      0.1 & $1.76 \pm 0.01$  &        0.7 & $-0.25 \pm 0.01$ &        4.76 \\
 NGC\ 6235     & 253.356  & -22.1774  & $11.94 \pm 0.38$ &      3.1 & $126.68 \pm 0.33$  &      0.3 & $-3.94 \pm 0.03$ &        0.8 & $-7.60 \pm 0.03$ &        0.41 \\
 NGC\ 6284     & 256.12   & -24.7648  & $14.21 \pm 0.42$ &      2.9 & $28.62 \pm 0.73$   &      2.5 & $-3.21 \pm 0.01$ &        0.4 & $-2.00 \pm 0.01$ &        0.7  \\
 NGC\ 6426     & 266.228  &   3.17014 & $20.71 \pm 0.35$ &      1.7 & $-210.51 \pm 0.51$ &      0.2 & $-1.82 \pm 0.02$ &        0.9 & $-3.00 \pm 0.01$ &        0.5  \\
 NGC\ 6584     & 274.657  & -52.2158  & $13.61 \pm 0.17$ &      1.3 & $260.64 \pm 1.58$  &      0.6 & $-0.08 \pm 0.01$ &       10    & $-7.20 \pm 0.01$ &        0.11 \\
 NGC\ 6981     & 313.365  & -12.5373  & $16.66 \pm 0.18$ &      1.1 & $-331.39 \pm 1.47$ &      0.4 & $-1.26 \pm 0.01$ &        0.6 & $-3.37 \pm 0.01$ &        0.15 \\
 NGC\ 7078     & 322.493  &  12.167   & $10.71 \pm 0.10$ &      0.9 & $-106.84 \pm 0.30$ &      0.3 & $-0.65 \pm 0.01$ &        1.4 & $-3.80 \pm 0.01$ &        0.21 \\
 ESO\ SC06     & 272.275  & -46.4233  & $20.95 \pm 0.65$ &      3.1 & $93.20 \pm 0.34$   &      0.4 & $-0.69 \pm 0.02$ &        2.8 & $-2.78 \pm 0.02$ &        0.58 \\
 Palomar\ 2        &  71.5246 &  31.3815  & $26.17 \pm 1.28$ &      4.8 & $-135.97 \pm 1.55$ &      1.1 & $1.23 \pm 0.05$  &        4.4 & $-1.53 \pm 0.04$ &        2.8  \\
 Djorgovski\ 1       & 266.87   & -33.0664  & $9.88 \pm 0.65$  &      6.6 & $-359.18 \pm 1.64$ &      0.5 & $-4.72 \pm 0.03$ &        0.6 & $-8.47 \pm 0.02$ &        0.25 \\
 Terzan\ 10       & 270.741  & -26.0669  & $10.21 \pm 0.40$ &      3.9 & $211.37 \pm 2.27$  &      1.1 & $-6.81 \pm 0.03$ &        0.4 & $-2.60 \pm 0.02$ &        0.89 \\
 Palomar\ 15       & 254.963  &  -0.539   & $44.10 \pm 1.14$ &      2.6 & $72.27 \pm 1.74$   &      2.4 & $-0.60 \pm 0.04$ &        5.8 & $-0.85 \pm 0.03$ &        3.39 \\
\hline
\end{tabular}}
\tablefoot{
Columns (1) -- the GCs names; (2)--(3) -- equatorial co-ordinates; (4)--(5) -- distance from the Sun and relative error; (6)--(7) -- radial velocity and corresponding relative error; (8)--(9) -- proper motion of $\alpha$ and relative error; column (10)--(11) -- proper motion of $\delta$ and relative error respectively.
}
\vspace{6pt}
\end{table*}

\begin{figure*}[h]
\centering
\includegraphics[width=0.89\linewidth]{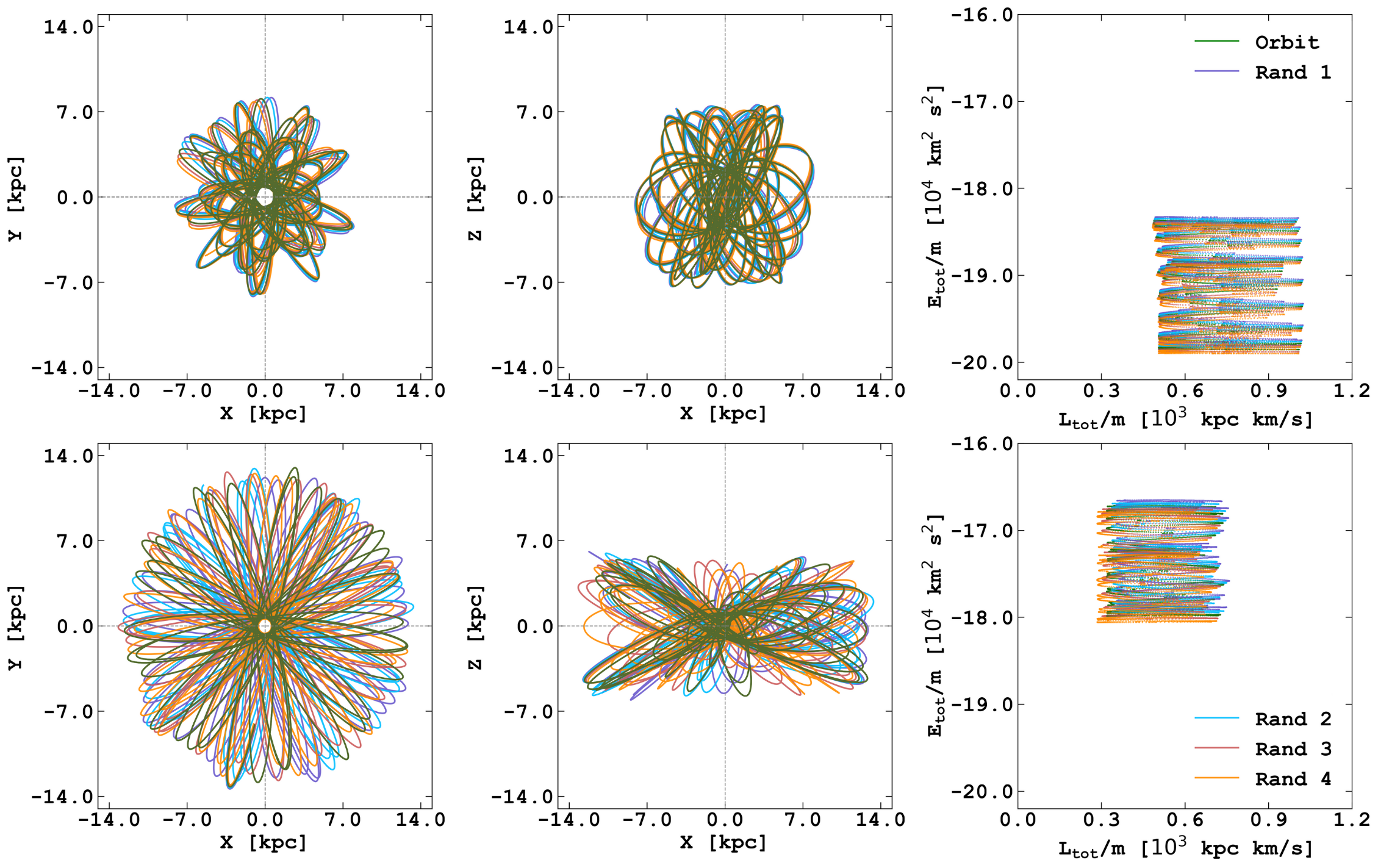}
\caption{Evolution of the GC orbits during a 5 Gyr lookback integration for NGC 7099 and NGC 5286. The left and centre panels show the orbits of GCs in \textit{X-Y} and \textit{X-Z} projections. Right panels show the GCs' evolution in terms of specific energy (E) with angular momentum L$_{\rm tot}$. The most probable orbit (middle values from \citealt{Baumgardt2021}) marked as green line -- `Orbit'. Other orbits marked as `rand' are random seeds in initial conditions for R$_{\odot}$, R$_{V}$, $\mu_{\alpha}$, and $\mu_{\delta}$, which are presented in Table~\ref{tab:cur-pos}.}
\label{fig:err-GC-orb}
\end{figure*}

To evaluate the influence of measurement errors on orbit shape during reconstruction, we generated four additional initial conditions for GCs taking into account the errors for parameters R$_\odot$, R$_V$, $\mu_\alpha$, and $\mu_\delta$ within $\pm$$\sigma$. For analysis we choose two extremes. One system, NGC 7099, is an example with a poor measurement accuracy and one system, NGC 5286, is a good example with a high measurement accuracy. In Fig.~\ref{fig:err-GC-orb} we show the result of the orbital evolution for both systems. As can be seen, the shape of the orbits is preserved and does not have a large deviations. Also, in the right panels of the Fig.~\ref{fig:err-GC-orb} we demonstrate the influence of measurement errors for the dynamics of total energy and angular momentum for selected GCs. As can be seen from Fig.~\ref{fig:err-GC-orb}, even for the GC's with the high relative errors in a particular component, this does not lead to a strong variation of the orbital turns over long intervals of backward integration in time.  

\clearpage

\section{Additional selection parameters}\label{sec:a-e-tot}

In Table~\ref{tab:select-criter} we provide a detailed list of our GC sample, organised in the same way as in Table~\ref{tab:gc-pre-sets}. The number of articles in which the GC satisfies the GE/S limits (as assumed by every author) in different kinematic parameters is shown in the columns after the slash. In the last row of the Table labelled `Chem.', we show the number of articles in which GCs were considered as GE/S progenitors according to chemical abundances and the age-metallicity relation~\cite{CARMA2025b, Callingham2022, Malhan2022, Kruijssen2020, Massari2019}, and ~\cite{Massari2019}.

\begin{table}[htbp]
\caption{Detailed estimation of the GE/S GCs.}
\centering
\sisetup{separate-uncertainty}
\begin{tabular}{@{}l@{}l@{}}
\begin{tabular}[t]{cccccccc}
\hline
\hline
GC's name & {\tt ecc} & r$_{\rm peri}$ & r$_{\rm apo}$ & L$_z$ & L$_{\rm perp}$ & E$_{\rm tot}$ & Chem.\\
\hline
\hline
 NGC\ 1261     & 3/3    & 1/1    & 2/2 & 4/4 & 2/2 & 3/4 & 5/5\\
 NGC\ 1851     & 3/3    & 1/1    & 2/2 & 4/4 & 2/2 & 3/4 & 5/5\\
 NGC\ 7089     & 3/3    & 1/1    & 2/2 & 4/4 & 2/2 & 3/4 & 5/5\\
 NGC\ 2808     & 3/3    & -/1    & 1/2 & 3/4 & 2/2 & 2/4 & 4/4\\
 NGC\ 2298     & 3/3    &  1/1   & 2/2 & 4/4 & 2/2 & 2/4 & 5/5\\
 NGC\ 6864     & 3/3    & 1/1    & 2/2 & 3/4 & 2/2 & 3/4 & 4/4\\
 NGC\ 6779     & 1/3    & 1/1    & 1/2 & 4/4 & 2/2 & 2/4 & 4/5\\
 NGC\ 4147     & 3/3    & 1/1    & 2/2 & 4/4 & 2/2 & 4/4 & 3/4\\
 NGC\ 5634     & 1/3    & 1/1    & 2/2 & 3/4 & 2/2 & 4/4 & 3/4\\
 NGC\ 5904     & 3/3    & 1/1    & 1/2 & 3/4 & 2/2 & 4/4 & 3/4\\
 NGC\ 7492     & 3/3    & 1/1    & 1/2 & 4/4 & 2/2 & 4/4 & 3/4\\
 NGC\ 1904     & 3/3    & -/1    & 2/2 & 4/4 & 2/2 & 3/4 & 3/4\\
 IC\ 1257      & 3/3    & -/1    & 2/2 & 4/4 & 2/2 & 3/4 & 3/3\\
 NGC\ 6229     & 3/3    & 1/1    & 1/2 & 3/4 & 2/2 & 4/4 & 2/3\\
 NGC\ 6981     & 3/3    & 1/1    & 2/2 & 4/4 & 2/2 & 3/4 & 2/4\\
\hline
 NGC\ 362      & 3/3    & -/1    & 1/2 & 4/4  & 2/2 & 1/4 & 4/5\\
 NGC\ 5286     & 3/3    & -/1    & 1/2 & 4/4  & 2/2 & 1/4 & 4/5\\
 NGC\ 6341     & 3/3 & -/1 & 1/2 & 4/4 & 2/2 & 1/4 & 4/5\\
\hline
\hline

\end{tabular}
&
\begin{tabular}[t]{|cccccccc}
\hline
\hline
GC's name & {\tt ecc} & r$_{\rm peri}$ & r$_{\rm apo}$ & L$_z$ & L$_{\rm perp}$ & E$_{\rm tot}$ & Chem.\\
\hline
\hline
 NGC\ 5897     & -/3  & 1/1 & 1/2 & 3/4    & 2/2  & 1/4 & 4/4 \\
 NGC\ 7099     & -/3  & 1/1 & 1/2 & 4/4    & 2/2  & -/4 & 4/5 \\
 NGC\ 6584     & 3/3    & 1/1     & 2/2 & 1/4 & 2/2 & 3/4 & 2/3\\
 ESO\ SC06     & 1/3 & 1/1 & 1/2  & 3/4 & 2/2 & 1/4 & 2/2\\
 Djorgovski\ 1 & 3/3 & 1/1 & 1/2  & 2/4 & 2/2 & 1/4 & 1/1\\
 NGC\ 6426     & -/3 & -/1 & 2/2  & -/4 & 2/2 & 3/4 & -/2\\
\hline
 NGC\ 288      & -/3  & 1/1 & 1/2 & 4/4    & 2/2  & 1/4 & 3/5 \\
 NGC\ 6205     & -/3  & 1/1 & 1/2 & 4/4    & 2/2  & -/4 & 3/5 \\
 NGC\ 6101     & -/3  & -/1 & -/2 & -/4    & 1/2  & 4/4 & 2/3 \\
 NGC\ 3201     & -/3  & -/1 & -/2 & -/4    & 2/2  & 3/4 & 2/3 \\
 NGC\ 6235     & -/3  & -/1 & 1/2 & -/4    & 2/2  & 1/4 & 3/3 \\
 NGC\ 5139     & -/3  & 1/1 & 1/2 & 3/4    & 2/2  & -/4 & 3/5 \\
 NGC\ 4833     & 3/3  & -/1 & 1/2 & 3/4    & 2/2  & -/4 & 2/3 \\
 Palomar\ 2    & 3/3  & -/1 & -/2 & 4/4    & 2/2  & 4/4 & 1/2 \\
 Palomar\ 15   & 3/3  & -/1 & -/2 & 4/4    & 2/2  & 3/4 & 1/2 \\
 NGC\ 7078     & -/3  & -/1 & 1/2 & -/4    & 2/2  & 1/4 & 1/3 \\
 NGC\ 6284     & 1/3  & -/1 & 1/2 & 4/4    & 2/2  & -/4 & 2/3 \\
 Terzan\ 10    & -/3  & 1/1 & 1/2 & 3/4    & 2/2  & -/4 & 1/2 \\
\hline
\hline
\end{tabular}
\end{tabular}
\label{tab:select-criter}
\end{table} 

\section{Estimation of relaxation time for the GCs with different particle numbers}\label{sec:time-relax}

\begin{figure*}[htbp!]
\centering
\includegraphics[width=0.99\linewidth]{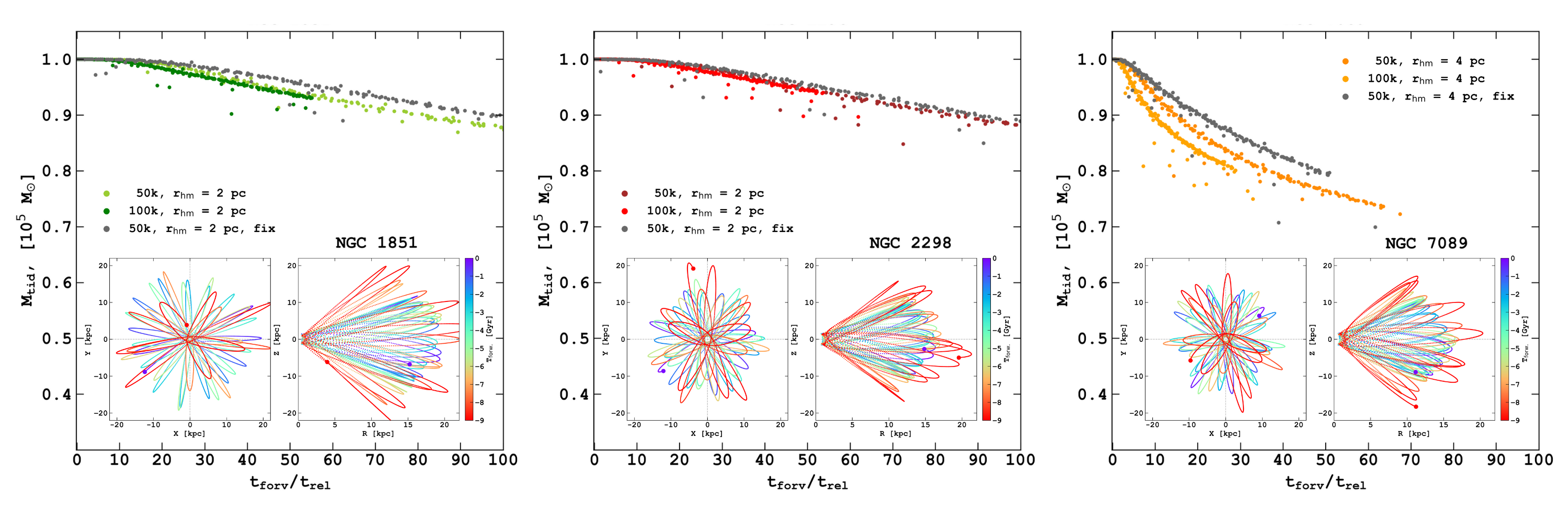}
\caption{Evolution of a GC's tidal mass depends on the ratio of its evolution time to its current relaxation time. The included graph also shows the orbits of a selected GC over 9 Gyr of evolution.}
\label{fig:t-relax-iniv-gs}
\end{figure*}

To investigate the dependence of star loss on particle number, we performed $N$-body simulations using a set of runs with larger number of particles (doubling the N). So, we created a new GC models with a particle number of N = 100\textit{k} with the same half-mass radius. We selected three GCs that are most strongly associated with GE/S: NGC 1851, NGC 2298, and NGC 7089. To calculate the current relaxation time, we use the following equation \citep{Binney2008}:
\begin{equation}
    t_{rel}=\frac{780 [{\rm Myr}]}{\ln{(\lambda \cdot N)}} \left(\frac{M}{10^5M_{\odot}}\right)^{1/2} \left(\frac{r_{hm}}{[{\rm pc}]}\right)^{3/2} \left(\frac{M_{\odot}}{m}\right)
    \label{eq:trel}
,\end{equation}where $m$ -- average mass of the single stellar particle, $M$ -- total current mass of the cluster, $N$ -- number of particles, r$_{hm}$ -- half-mass radius; $\lambda$ = 0.1 value we also took from \cite{Giersz1994}. Figure~\ref{fig:t-relax-iniv-gs} shows the evolution of the tidal mass of the clusters as a function of the normalised time (i.e. evolution time divided by relaxation time). The relaxation time is calculated at every 10 Myr over a whole computation period. As can be seen for NGC 1851, NGC 2298, and NGC 7089, the mass evolution does not depend on the number of particles in our models. Models, shown with grey curves, represents the GC's stellar loss in another realisation of the galactic potential; see more details in Appendix \ref{sec:test-runs}.

\section{Some additional test simulations with different conditions}\label{sec:test-runs}

\subsection{Evolution of the orbital parameters in the {\tt 441327} TVP potential}\label{subsec:check-e-ltot-441327}

\begin{figure*}[htbp!]
\centering
\includegraphics[width=0.90\linewidth]{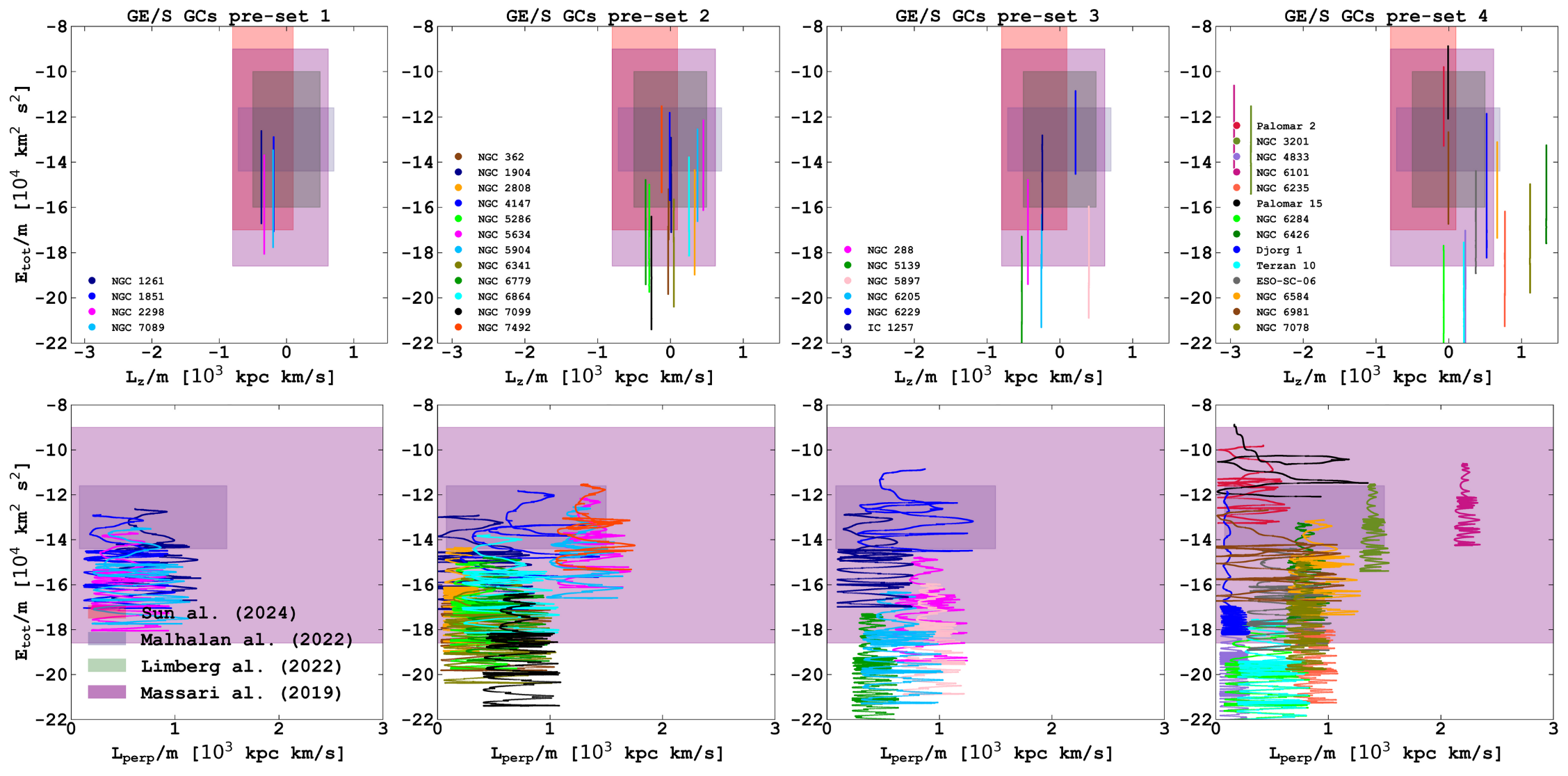}
\caption{Evolution of the GCs phase-space parameters during the 9 Gyr lookback time integration in {\tt 441327} TVP external potential for the 4 different pre-sets of GE/S GCs.}
\label{fig:E-lz-sets-441327}
\end{figure*}

\begin{figure*}[htbp!]
\centering
\includegraphics[width=0.90\linewidth]{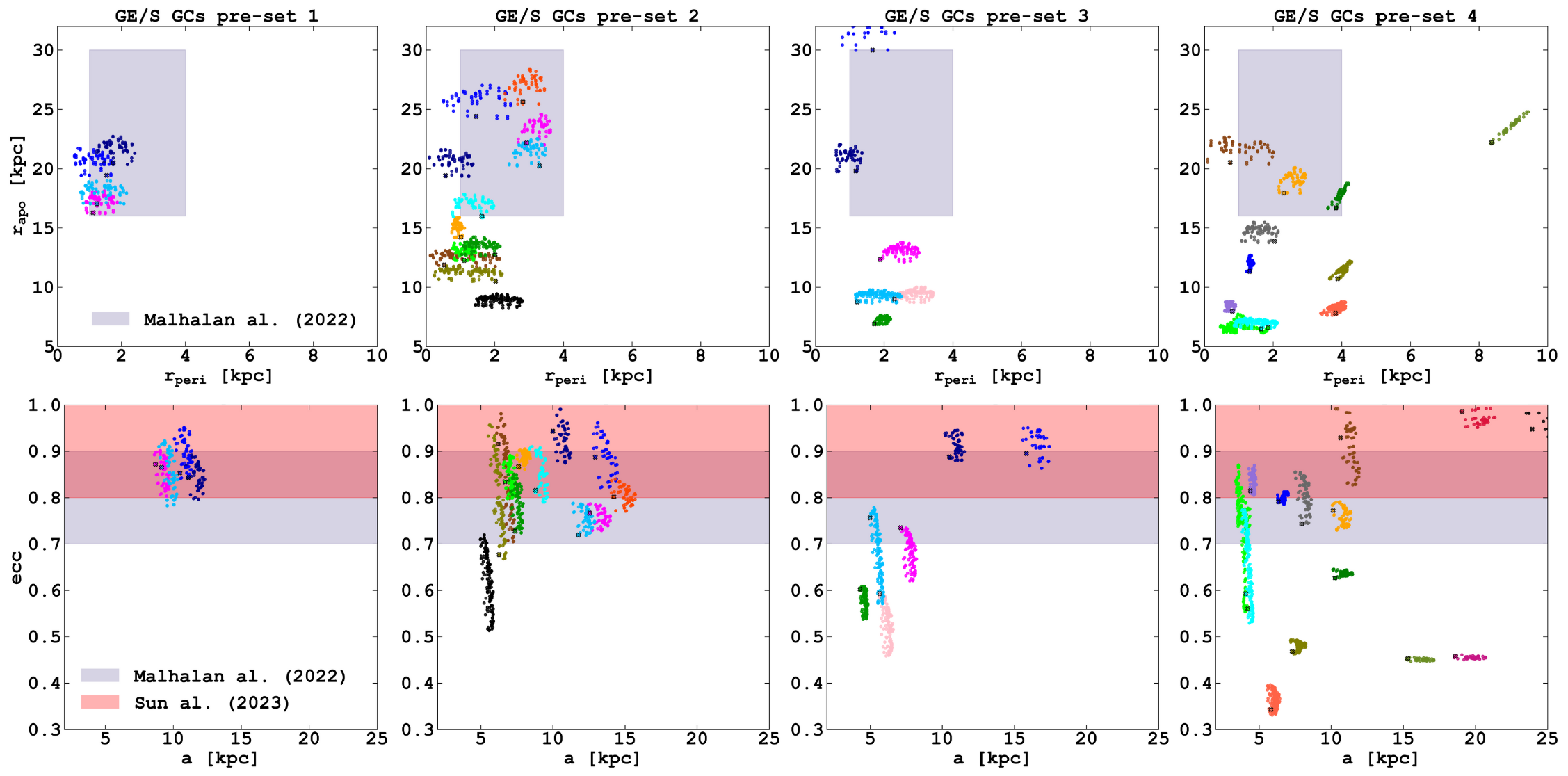}
\caption{Evolution of the GCs orbital parameters during the 9 Gyr lookback time integration in {\tt 441327} TVP external potential for 4 different pre-sets of GE/S GCs.}
\label{fig:a-e-sets-441327}
\end{figure*}

As stated in Sect.~\ref{subsec:res-sam}, to make our result more robust, we investigate the energy vs. angular momentum evolution together with the orbital elements changes in one more external TVP. We selected {\tt 441327} external time-variable MW-like potential to determine the possible dependency of the resulting GE/S GC sample from the external potential. We carried out a point-mass integration of 36 GCs in the same way as we described in Sect. ~\ref{sec:orb-reconstr} but in the {\tt441327} TVP potential. In Fig.~\ref{fig:E-lz-sets-441327} we show the evolution of the total specific energy and the angular momentum of L$_z$ and L$_{perp}$ for individual GCs. In Fig.~\ref{fig:a-e-sets-441327} we also present the evolution of the orbital parameters for these GCs. 

\subsection{Evolution of the orbital parameters in the time-independent potential based on {\tt 411321} TVP parameters}\label{subsec:tvp-fix}

\begin{figure*}[htbp!]
\centering
\includegraphics[width=0.95\linewidth]{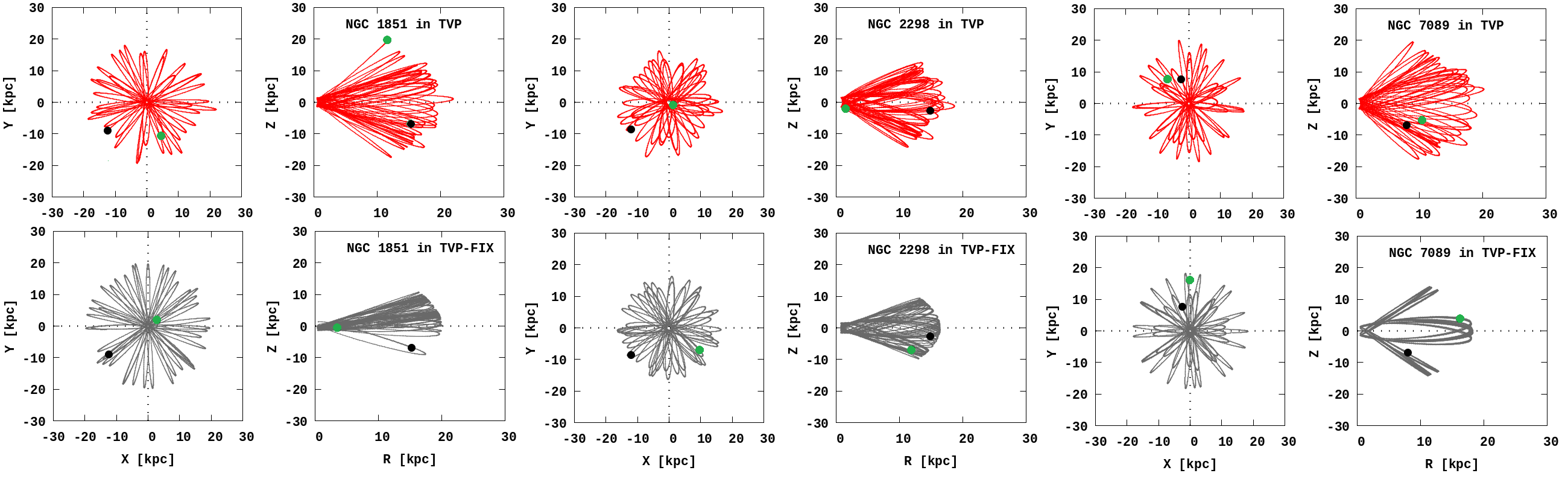}
\caption{Orbital evolution in the {\tt 411321} TVP external potential (red colour) and in time-independent potential (gray colour), presented in $X-Y$ an $R-Z$, where $R$ is the planar Galactocentric radius. The total time of integration is a 9 Gyr lookback time. Black circle shows current position,  green -- at -9 Gyr ago.}
\label{fig:dccorr-tng-fix}
\end{figure*}

For our additional simulations we analyse mass loss from GCs with another realisation of the time-independent potential, {\tt 411321 TVP-FIX}. We take the current masses and scale parameters for 411321 TVP as it is today. For disc we assume the following values: $M_{\rm d}$ = 7.110$\times10^{10}~\rm M_{\odot}$; scale length $a_{\rm d}$ = 2.073~kpc; scale height $b_{\rm d}$ = 1.126~kpc. For halo: $M_{\rm h}$ = 1.190$\times10^{12}~\rm M_{\odot}$; scale height $b_{\rm h}$ = 28.48 kpc; see also  Fig. \ref{fig:ext-pot} in Sect. \ref{sec:orb-reconstr}. We fixed these values of external potential parameters and carried out the $N$-body orbital reconstruction up to -9 Gyr for the three GCs as a single particle: NGC 1851, NGC 2298, and NGC 7089. For comparison of the potential influence on the orbital reconstruction, we show two types of simulations: red represents the orbital evolution in TVP potential, and gray in TVP-FIX; see Fig. \ref{fig:dccorr-tng-fix}. As we see from these plots, applying these two different potentials, we get almost similar orbital shapes only with some minor differences, which. In our opinion, these changes do not generally have a strong influence on the orbits of these GCs. Of course, the TVP with time-dependent parameters is more physically motivated compared to the TVP-FIX for the long term dynamical simulations of GC dynamics in MW.

Also, to investigate the dependence from selected TVP and TVP-FIX we check the dependence from particle number in the models (NGC 1851, NGC 2298, and NGC 7089). For these clusters, we created completely new models with a particle number $N$ = 50\textit{k}, the same half-mass radii and integrated them forward for 9 Gyr in TVP-FIX. As we see in Fig. \ref{fig:t-relax-iniv-gs}, for clusters NGC 1851 and NGC 2298, the models that evolve in TVP and TVP-FIX, variations in mass loss are almost the same. The reason for such a behaviour is probably related to the compactness of the cluster models with r$_{\rm hm}$ = 2 pc. The NGC 7089 model with a larger r$_{\rm hm}$ = 4 pc has a slightly different mass loss, but during the entire simulation the difference between the TVP and TVP-FIX mass loss results is not greater than a few percentage points.

\subsection{Evolution of the orbital parameters in the {\tt 41321} TVP potential with dynamical friction included}\label{subsec:dyn-fr}

In our current work, we apply additional calculations to determine the influence of dynamical friction on our orbital parameter analysis for the GCs. Dynamical friction in astrophysical context indicates the collective deceleration exerted on a moving massive body by the fluctuating force of field stars. The existence of such an effect was first demonstrated by Chandrasekhar and von Neumann \citep{CN1942, CN1943} in their pioneering works. Later, Chandrasekhar in the studies of \citep{C1943a, C1943b} developed a more quantitative theory of dynamical friction. The resulting drag acceleration acting on the GC can be written as \citep{Binney2008} 
\begin{equation}
\frac{{\rm d} {\bf V}_{\rm GC} }{\rm dt} = -\frac{4\pi G^2\rho M_{\rm GC}}{V_{\rm GC}^3} \; \chi \cdot 
        \ln\Lambda \cdot {\bf V}_{\rm GC} 
    \quad \mathrm{with}\quad 
    \chi=\frac{\rho(<V_{\rm GC})}{\rho}.
    \label{dynfric}
\end{equation}

In general, the functions $\chi$ and the Coulomb logarithm $\Lambda$ depends on the velocity of the massive object and the properties of the background system. For more details on this dynamical friction description, we refer the reader to \cite{Just2011}. Based on the results of our previous study \cite{Just2011}, in our current calculation we use the fixed Coulomb logarithm value $\ln\Lambda$ = 5 and the fixed value for $\chi$ = 0.5. 

\begin{figure*}[htbp!]
\centering
\includegraphics[width=0.95\linewidth]{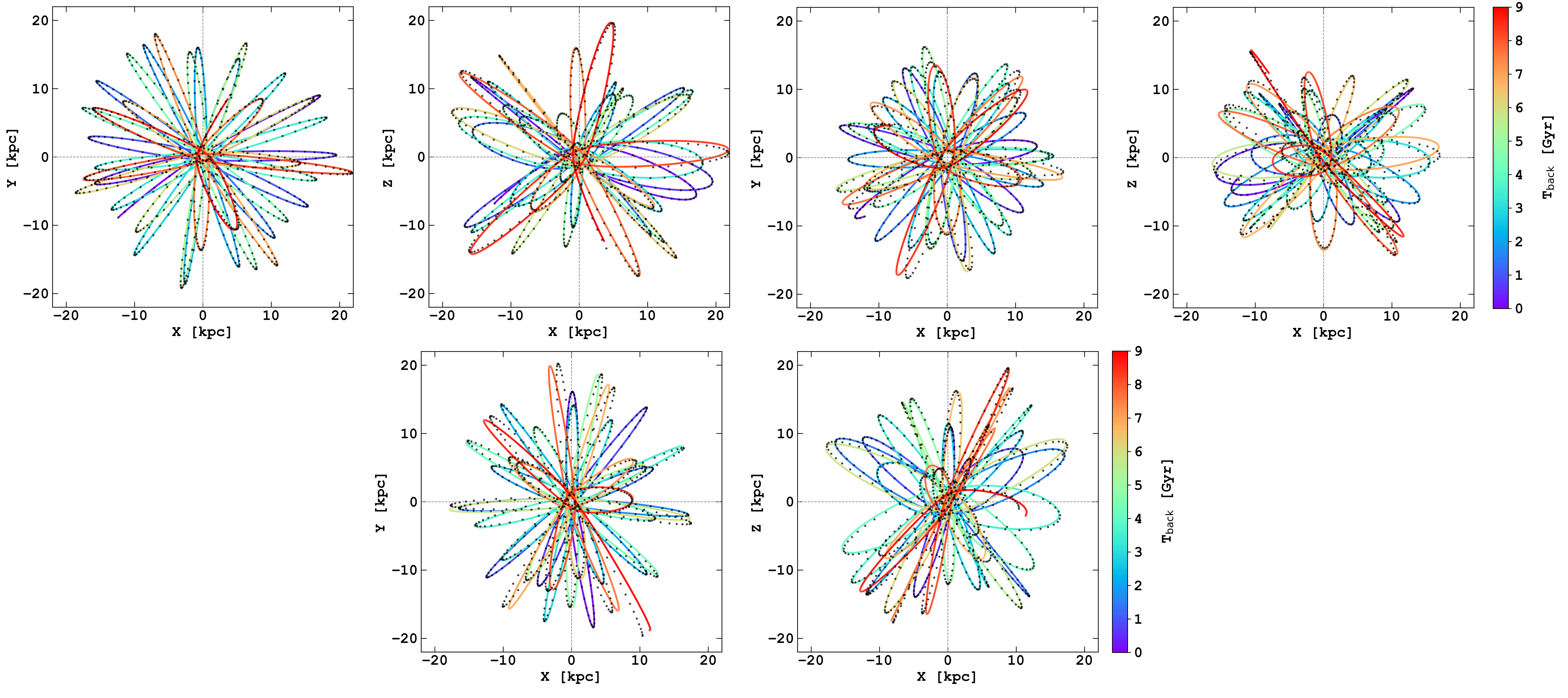}
\caption{Evolution of a GC’s Galactocentric co-ordinates in the {\tt 411321} TVP, colored curve, and in the same {\tt 411321} TVP but with the dynamical friction, black dots, based on the 9 Gyr backward integration.}
\label{fig:t-dyn}
\end{figure*}

In Fig. \ref{fig:t-dyn} we present the comparison of the evolution of the GC Galactic orbits during 9 Gyr of integration time for the models including dynamical friction and without it. As can be seen from the plot, dynamical friction only slightly affects on orbits. In our opinion, applying the dynamical friction in a present form does not significantly change the orbit of GC and so the position of lost stars from these GCs on the phase-space diagrams.

\end{appendix}
\end{document}